\documentstyle[12pt]{report}
\def\thebibliography#1{\nonumsection{\large \it References}\list
  {[\arabic{enumi}]}{\settowidth\labelwidth{[#1]}\leftmargin\labelwidth
    \advance\leftmargin\labelsep
    \usecounter{enumi}}
    \def\newblock{\hskip .11em plus .33em minus .07em}
    \sloppy\clubpenalty4000\widowpenalty4000}

\newcommand{\tcaption}[1]{                      
        \addtocounter{table}{1}
         {{\tenrm\offinterlineskip Table~\thetable . #1} }\hfil\break }

\newcommand{\nonumsection}[1] {\vspace{12pt}\noindent{\bf #1}
        \par\vspace{5pt}}

\newcommand{\be}{\begin{eqnarray}}
\newcommand{\ee}{\end{eqnarray}}
\newcommand{\dslash}{\partial \hskip -0.5em /}
\newcommand{\Dslash}{D \hskip -0.7em /}
\newcommand{\Gslash}{\Gamma \hskip -0.7em /}
\newcommand{\tr}{{\rm tr}}

\newcommand{\A}{{\cal A}}

\newcommand{\textlineskip}{\baselineskip=14pt}

 1
 1
 1

\begin{document}

{\thispagestyle{empty}}
\setcounter{page}{1}

\rightline{UNITU-THEP-12/1994}
\rightline{June 1994}
\rightline{hep-ph/9406420}

\vspace{1cm}
\centerline{\Large \bf Topologically non--trivial chiral
transformations:}
\vspace{0.5cm}
\centerline{\Large \bf The chiral invariant elimination of the
axial vector meson$^\dagger $}
\vspace{0.5cm}
\centerline{R.\ Alkofer, H.\ Reinhardt, J.\ Schlienz$^\ddagger$ and
H.\ Weigel$^\natural$}
\vspace{0.2cm}
\centerline{Institute for Theoretical Physics}
\vspace{0.2cm}
\centerline{T\"ubingen University}
\vspace{0.2cm}
\centerline{Auf der Morgenstelle 14}
\vspace{0.2cm}
\centerline{D-72076 T\"ubingen, FR Germany}

\vspace{2cm}
\centerline{\bf Abstract}
\vspace{0.5cm}
\baselineskip14pt
\noindent
The role of chiral transformations in effective theories
modeling Quantum Chromo Dynamics is reviewed. In the context
of the Nambu--Jona--Lasinio  model the hidden gauge and massive
Yang--Mills approaches to vector mesons are demonstrated to be
linked by a special chiral transformation which removes the chiral
field from the scalar--pseudoscalar sector. The chirally rotated
axial vector meson field ($\tilde A_\mu$) transforms homogeneously
under flavor rotations and may thus be dropped without violating
chiral symmetry. The fermion determinant for static meson field
configurations is computed by summing the discretized eigenvalues of
the Dirac Hamiltonian. It is discussed how the local chiral
transformation loses its unitary character in a finite model space.
This technical issue proves to be crucial for the construction of the
soliton within the Nambu--Jona--Lasinio  model when the chirally
rotated axial vector field is neglected. In the background of this
soliton the valence quark is strongly bound, and its eigenenergy turns
out to be negative. This important physical property which is usually
generated only by non--vanishing axial vector is thus
carried over by the simplification $\tilde A_\mu=0$.


\vfill
\noindent
$^\dagger $
{\footnotesize{Supported by the Deutsche Forschungsgemeinschaft (DFG)
under contract  Re 856/2-1.}}
\newline
$^\ddagger $
{\footnotesize{Present address: Institute for Theoretical Physics and
Synergetic, Stuttgart University,\newline
$\mbox{ }$ Pfaffenwaldring 57, D-70550
Stuttgart.}}
\newline
$^\natural$
{\footnotesize{Supported by a Habilitanden--scholarship of the DFG}}
\eject

\normalsize\textlineskip
\stepcounter{chapter}
\leftline{\large\it 1. Introduction}
\bigskip

Since a solution to Quantum Chromo Dynamics (QCD) is not yet
available one has to recede on models in order to explore processes
described by the strong interaction. These models are usually
constructed under the requirement that the symmetries of the
underlying theory, {\it i.e.} QCD are maintained. In this context
chiral symmetry and its spontaneous breaking are of special importance.
In this article we will explore a special chiral transformation
when topologically non--trivial meson field configurations like
solitons are involved. To begin with, let us briefly review the
relevance of chiral symmetry on the one side and solitonic field
configurations on the other in the context of strong interactions.

QCD can be extended from $SU(3)$ to $SU(N_C)$ where $N_C$
denotes the number of color degrees of freedom. It was observed
by `t Hooft \cite{tho74} that in the limit $N_C\rightarrow\infty$
QCD is equivalent to an effective theory of weakly interacting mesons.
Subsequently Witten \cite{wi79} conjectured that baryons emerge as
solitons of the meson fields within this effective theory. Stimulated
by Witten's conjecture much interest has been devoted to the
description of baryons as chiral solitons during the past
decade \cite{ad83,ho93}. In the soliton description of baryons the
chiral field and in particular its topological structure play a key
role. The topological character of the chiral field especially endows
the soliton with baryonic properties like baryon charge and
spin \cite{wi83}. This comes about via the chiral anomaly \cite{ba69}
which is a unique feature of all quantum field theories where fermions
live in a gauge group and couple to a chiral field. The fermionic part
of such a theory has the generic structure
\be
Z_F[\Phi]
&=& \int D \Psi D \bar{\Psi}\ \mbox{exp}\left[
i\int\ d^4x\ \bar{\Psi}\left(i \Dslash\ -\hat m_0\right)
\Psi\right]  \nonumber\\
&=& \mbox{Det}\, \left(i\Dslash\ -\hat m_0\right)
\label{zferm}
\ee
where $\hat m_0$ denotes the current quark mass matrix which will
be ignored in the ongoing discussions. Furthermore
\be
i\Dslash
= i \dslash- \Phi
= i(\dslash + \Gslash ) - MP_R - M^{\dag}  P_L
\label{Dslash}
\ee
represents the Dirac--operator of the fermions in the external
Bose--field $\Phi$. $\Phi$ in general contains vector $V_\mu$,
axial vector $A_\mu$ fields\footnote{In our notation $V_\mu$ and
$A_\mu$ are anti--hermitian.} as well as scalar $S$ and
pseudo-scalar fields $P$
\be
\Gamma_\mu = V_\mu + A_\mu\gamma_5,
\qquad M=S+iP=\xi_L^{\dag}  \Sigma \xi_R.
\label{vecfield}
\ee
Here $P_{R/L} =  \frac{1}{2}(1\pm\gamma_5)$ are the chiral
projectors. Accordingly one defines left(L)-- and
right(R)--handed quark fields: $\Psi_{L,R}=P_{L,R}\Psi$. The
chiral field $U$ is defined via the polar decomposition
of the meson fields
\be
U = \xi_L^{\dag}  \xi_R = {\rm exp}(i\Theta).
\label{chifield}
\ee

The chiral anomaly arises because there is no regularization
scheme which simultaneously preserves local vector and axial
vector (chiral) symmetries. In renormalizable theories the
chiral anomaly can be calculated in closed form and is given
by the Wess-Zumino action \cite{we79,wi83,kay84}
\be
{\cal A} = S_{WZ} = \frac{iN_C}{240\pi^2}
\int_{M^5}(UdU^{\dag} )^5
\label{wzw}
\ee
where the notation of alternating differential forms has been used.
$M^5$ denotes a five dimensional manifold whose boundary is
Minkowski space. Obviously the chiral anomaly is tightly related
to the chiral field since the Wess-Zumino action vanishes when
the chiral field disappears $(U = 1)$. The chiral anomaly is,
however, not merely a technical artifact but has well established
physical consequences.  In the meson sector it gives rise to the
so--called ``anomalous decay processes'' like {\it e.g.}
$\pi \rightarrow 2\gamma$ and $\omega \rightarrow 3\pi$. In the
soliton sector the chiral anomaly requires for $N_C = 3$ the soliton
to be quantized as a fermion and endows the soliton with half integer
spin and integer baryon number\footnote{For this proof it is
mandatory to consider flavor SU(3).} \cite{wi83,ma84}.

For many purposes it is convenient to perform a chiral rotation
of the fermions \cite{eb86,re89a}
\be
\tilde{\Psi} = \Omega \Psi\quad {\rm with} \qquad
\Omega = P_L \xi_L + P_R\xi_R,
\quad {\it i.e.}\quad \tilde{\Psi}_{L,R}=\xi_{L,R}\Psi_{L,R}.
\label{psitilde}
\ee
This transformation defines a chirally rotated Dirac-operator
\be
\Psi i \Dslash\,\Psi = \bar{\tilde{\Psi}} i \tilde{\Dslash} \, \tilde{\Psi}
\label{rot1}
\ee
which acquires the form
\be
i\tilde{\Dslash} = \Omega^{\dag}  i \Dslash \, \Omega^{\dag}  =
i\gamma_\mu\left(\partial^\mu+\tilde{V}^\mu
+\tilde{A}^\mu\gamma_5\right)-\Sigma.
\label{rot2}
\ee
The chiral rotation has removed the chiral field from the scalar
pseudo-scalar sector of the rotated Dirac operator $i\tilde{\Dslash}$.
As a consequence the vector and axial vector fields now become
chirally rotated
\be
\tilde{V}_\mu + \tilde{A}_\mu
&=& \xi_R(\partial_\mu + V_\mu + A_\mu) \xi_R^{\dag} , \nonumber\\
\tilde{V}_\mu - \tilde{A}_\mu
&=& \xi_L(\partial_\mu + V_\mu - A_\mu) \xi_L^{\dag} .
\label{rotvec}
\ee
Even in the absence of vector and axial vector
fields in the original Dirac operator $(V_\mu = A_\mu = 0)$
the chiral rotation induces vector and axial vector fields
\be
\tilde{V}_\mu (V_\mu = A_\mu = 0)=v_\mu
&=&\frac{1}{2}\left(\xi_R \partial_ \mu \xi_R^{\dag}  +
\xi_L \partial_\mu \xi_L^{\dag} \right),
    \nonumber\\
\tilde{A}_\mu (V_\mu = A_\mu = 0)=a_\mu
&=& \frac{1}{2}\left(\xi_R \partial_\mu \xi_R^{\dag}  -
\xi_L\partial_\mu \xi_L^{\dag} \right).
\label{indvec}
\ee

For the soliton description of baryons the chiral field is usually
assumed to be of the hedgehog type
\be
U = {\rm exp}\left(
i\Theta(r){\mbox{\boldmath $\tau$}}
\cdot{\hat{\mbox{\boldmath $r$}}}\right)
\label{hedgehog}
\ee
The non--trivial topological structure of this configuration is then
exhibited by the boundary conditions $\Theta(0)=-n\pi$ and
$\Theta(\infty)=0$. The chiral field thus represents a mapping
from the compactified coordinate space (all points at spatial infinity
are identified) to SU(2) flavor space, {\it i.e.} $S^3\rightarrow S^3$.
The associated homotopy group, $\Pi_3(S^3)$, is isomorphic to $Z$, the
group of integer numbers. The isomorphism is given by the winding number
$\left(\Theta(0)-\Theta(\infty)\right)/\pi=-n$. Assuming the unitary
gauge $(\xi_L^{\dag} =\xi_R)$  the induced vector field
is of the Wu--Yang form \cite{re89a}
\be
v_0 = 0, \quad
v_i = i v_i^a \frac{\tau^a}{2}, \quad
v_i^a = \epsilon^{ika} \hat{r}_k \frac{G(r)}{r}
\label{wuyang}
\ee
with the profile function $G(r)$ given by the chiral angle $\Theta(r)$
\be
G(r) = -2 {\rm sin}^2 \frac{\Theta(r)}{2}.
\label{gtheta}
\ee
For odd $n$ the topological non--trivial character of the chiral
rotation is also reflected by a non--vanishing value of the
induced vector field at $r=0$.
The induced axial vector field $a_i = i a_i^a \tau^a/2$
becomes
\be
a_i^a =\hat{r}_i \hat{r}_a
\left(\Theta^\prime(r) - \frac{\sin\Theta(r)}{r}\right)
+ \delta_{ia} \frac{\sin\Theta(r)}{r}
\label{atheta}
\ee
where the prime indicates the derivative with respect to the argument.

The use of the chirally rotated fermions is advantageous since the
rotated Dirac-operator does no longer contain a chiral field and
its determinant is hence anomaly free. The chiral anomaly,
however, has not been lost by the chiral rotation but is hidden
in the integration measure over the fermion fields. In fact, as
observed by Fujikawa \cite{fu80}, a chiral rotation of the fermion
fields gives rise to a non--trivial Jacobian of the integration
measure
\be
D\Psi D\bar{\Psi} = J(U) D \tilde{\Psi} D \bar{\tilde{\Psi}}
\label{measure}
\ee
which is precisely given by the chiral anomaly
\be
J(U) = \mbox{exp}\,\left(i\cal{A}\right)
\label{measano}
\ee
Thus we have the relation
\be
- i {\rm Tr} \log i \Dslash = - i {\rm Tr} \log i\tilde{\Dslash} + \cal{A}
\label{actiontrans}
\ee
In many cases it is convenient to work with the chirally rotated
fermion fields because of the absence of the anomaly from the
fermion determinant.

As already mentioned above the chiral anomaly or equivalently the
non--trivial Jacobian in the fermionic integration measure arise
due to the need for regularization, which introduces a finite cut--off.
In the regularized theory the chiral anomaly can be evaluated in a
gradient expansion. In leading order the anomaly is then given by
the Wess-Zumino action (\ref{wzw}). There are, however, higher order
terms which are suppressed by inverse powers of the cut--off. In
renormalizable theories where the cut--off goes to infinity, these
higher order terms disappear and the chiral anomaly is known in closed
form. In non--renormalizable effective theories, however, the cut--off
of the regularization scheme has to be kept finite and acquires a
physical meaning, indicating the range of validity of the effective
theory. In this case the higher order terms of the gradient expansion
do not disappear but contribute to the anomaly which is then no longer
available in closed form.

Furthermore, when the soliton sector of such effective mesonic
theories is studied it is not sufficient to only consider the leading
and sub--leading contributions from the gradient expansion and one
has to perform a full non--perturbative evaluation of the fermion
determinant \cite{re89}. The non--perturbative calculations have to
be performed numerical\-ly \cite{re88}, with the continuous space being
discretized. This is the case for both, coordinate and momentum space.
Also in the non--perturbative studies of the soliton
sector of the effective theory the use of the chiral rotation is in
many cases advantageous \cite{re89a}. As noticed above for the soliton
description of baryons the topological nature of the chiral field is
crucial. Actually, topology is a property of continuous spaces
(manifolds) and it is {\it a priori} not clear whether the chiral
rotation with a topologically non--trivial chiral field can be
represented in a finite dimensional and discretized model space used
in the numerical calculations. In this respect let us recall that a
single point defect in a manifold changes its topological properties
already drastically.

The purpose of this paper is twofold. First we wish to present a study
of the chiral rotation in non--perturbative soliton calculations where
the fermion determinant has to be numerically evaluated in the
background of a topologically non--trivial chiral field. It will be
demonstrated how this rotation influences the choice of the boundary
conditions for the eigenstates of the Dirac Hamiltonian. Second we
will make use of the fact that in the chirally rotated formulation
the axial vector field may be set to zero ({\it i.e.} $\tilde A_\mu=0$)
without spoiling chiral symmetry. The resulting soliton configuration
is constructed and compared with various solutions in the unrotated
formulation. For definiteness we will use the Nambu-Jona-Lasinio
(NJL) \cite{na61} model as a microscopic fermion theory which shares
all the relevant properties of chiral dynamics with QCD. Its
bosonized version gives a quite satisfactory description of
mesons \cite{eb86} and also of baryons when the soliton picture is
assumed \cite{al93}.

The organization of the paper is as follows: After these introductory
remarks we review the importance of the local chiral rotation
(\ref{psitilde}) for the extraction of meson properties from the
NJL action. In that section also the chirally invariant elimination
of the axial vector meson is described. In section 3 we review the
computation of the soliton solution to the NJL model of pseudoscalar
mesons. This may straightforwardly be generalized to the inclusion
of other mesons as long as the Euclidean Dirac Hamiltonian remains
Hermitian. Section 4 is devoted to the study of the local chiral
rotation for topologically non--trivial field configurations and
its influence on the soliton solution. The computation of the
soliton solution with the rotated axial vector meson field being
eliminated is described in section 5. Previously it has been shown
that the presence of the axial vector field is responsible for a
strong binding of the valence quark and the valence quarks' eigenenergy
being negative \cite{al92,do92}. We will in particular examine whether
this important piece of information is maintained after eliminating
$\tilde A_\mu$. This is not obvious because simply setting
$A_\mu=0$ leads to a positive valence quark energy for a reasonable
choice of parameters \cite{al90}. A concluding discussion is given
in section 6. Some technical remarks on the boundary conditions
for the Dirac spinors are left as an appendix.

\bigskip
\stepcounter{chapter}
\leftline{\large\it 2. Properties of local chiral transformations}
\bigskip

As indicated in the introduction the chirally rotated formulation
of the NJL model is suited to investigate properties of (axial--)
vector mesons. In the present section we will therefore briefly
review the results and illuminate the connection with the
hidden gauge symmetry (HGS) and massive Yang--Mills (MYM) approaches
for the description of (axial--) vector mesons. Many of the
results reported in this section are taken from earlier
works \cite{eb86,wa88,re89a,wa89}. Nevertheless we repeat these
results here in order to put our work into perspective and have the
paper self--contained.

In the chirally rotated formulation the bosonized version of the
NJL model action reads
\be
{\cal A}_{\rm NJL}=- i {\rm Tr} \log i\tilde{\Dslash} + {\cal A}
-\frac{1}{4G_1}{\rm tr} \left(\Sigma^2 - m^2 \right)
-\frac{1}{4G_2}{\rm tr} \left[\left(\tilde V_\mu-v_\mu\right)^2
+\left(\tilde A_\mu-a_\mu\right)^2\right]
\label{njlrot}
\ee
with the constituent quark mass $m$ being the vacuum expectation value
of the scalar field $\Sigma$ {\it i.e.} $\langle\Sigma\rangle=m$. Again
we have discarded terms proportional to the current quark mass matrix.
The chirally transformed (axial--) vector fields are defined in eqns
(\ref{rotvec}) and (\ref{indvec}). Next we have to face the fact
that the functional trace of the logarithm in (\ref{njlrot}) is
ultra--violet divergent and thus needs regularization. This is
achieved by first continuing to Euclidean space $(x_0=-ix_4)$ and then
representing the real part of the Euclidean action by a parameter
integral
\be
\frac{1}{2}{\rm Tr} \log \left( \tilde{\Dslash}_E^{\dag}
\tilde{\Dslash}_E\right) \longrightarrow
-\frac{1}{2}\int_{1/\Lambda^2}^\infty \frac{ds}{s}
{\rm exp}\left(-s \tilde{\Dslash}_E^{\dag}
\tilde{\Dslash}_E\right)
\label{regreal}
\ee
which introduces the cut--off $\Lambda$. This substitution is an
identity up to an irrelevant additive constant for
$\Lambda\rightarrow\infty$. The Euclidean Dirac operator
$\tilde{\Dslash}_E$ is obtained by analytical continuation of
$\tilde{\Dslash}$ to Euclidean space. The prescription
(\ref{regreal}) is known as the proper--time
regularization \cite{sch51}. For the purpose of the present paper
it is sufficient to only consider the normal parts of the action.
We may therefore neglect the imaginary part as well as the anomaly
${\cal A}$. Thus the actual starting point of our considerations is
represented by
\be
{\cal A}_{\rm NJL}&=&-\frac{1}{2}\int_{1/\Lambda^2}^\infty \frac{ds}{s}
{\rm Tr}\ {\rm exp}\left(-s \tilde{\Dslash}_E^{\dag}
\tilde{\Dslash}_E\right)
-\frac{1}{4G_1}{\rm tr} \left(\Sigma^2 - m^2 \right)
\nonumber \\ &&\hspace{3cm}
-\frac{1}{4G_2}{\rm tr} \left[\left(\tilde V_\mu-v_\mu\right)^2
+\left(\tilde A_\mu-a_\mu\right)^2\right].
\label{stpoint}
\ee
In order to extract information about the properties of the (axial--)
vector mesons commonly a (covariant) derivative expansion of the
fermion determinant is performed. We choose to consider a covariant
derivative expansion since, in contrast to an on--shell determination
of the parameters \cite{ja92}, it preserves gauge invariance and does
not lead to artificial mass terms. Furthermore, this procedure leaves
the extraction of the axial--vector meson mass unique. Continuing back
to Minkowski space and substituting the scalar field $\Sigma$ by its
vacuum expectation value yields the leading terms \cite{wa88,re89a}
\be
{\cal L}_{\rm NJL}=\frac{1}{2g_V^2}{\rm tr}\
\left(\tilde V_{\mu\nu}^2+\tilde A_{\mu\nu}^2\right)
-\frac{6m^2}{g_V^2}{\rm tr}\ \tilde A_\mu^2
-\frac{1}{4G_2}{\rm tr} \left[\left(\tilde V_\mu-v_\mu\right)^2
+\left(\tilde A_\mu-a_\mu\right)^2\right] + \ldots\ .
\label{expand}
\ee
Here $\tilde V_{\mu\nu}$ and $\tilde A_{\mu\nu}$ denote the field
strength tensors
\be
\tilde V_{\mu \nu }&=& \partial_\mu\tilde V_\nu-\partial_\nu \tilde
V_\mu+[\tilde V_\mu, \tilde V_\nu]+[\tilde A_\mu, \tilde A_\nu],
\nonumber \\*
\tilde A_{\mu \nu} &=& \partial_\mu \tilde A_\nu - \partial_\nu
\tilde A_\mu+[\tilde V_\mu, \tilde A_\nu]+[\tilde A_\mu, \tilde V_\nu]
\label{ftensor}
\ee
of chirally rotated vector and axial--vector fields, respectively.
Obviously in our convention these fields contain the coupling constant
$g_V$ which in the proper--time regularization is given by
\be
g_V=4\pi\left[\frac{2N_C}{3}
\Gamma\left(0,\left(\frac{m}{\Lambda}\right)^2\right)\right]
^{-\frac{1}{2}}.
\label{copconst}
\ee
For the description of the pion fields we adopt the unitary
gauge for the chiral fields: $\xi=\xi^{\dag} _L=\xi_R$. The
pions come into the game by the non--linear realization
$\xi={\rm exp}\left(i\mbox{\boldmath $\tau$}\cdot\mbox{\boldmath
$\pi$}/f\right)$. Then the last term in eqn (\ref{expand}) contains
the axial--vector pion mixing which is eliminated by a corresponding
shift in the axial field: $\tilde A_\mu\rightarrow\tilde A^\prime_\mu
=\tilde A_\mu+(ig_V^2m^2/12fG_2)\partial_\mu
\mbox{\boldmath $\tau$}\cdot\mbox{\boldmath $\pi$}$. This shift
obviously provides an additional kinetic term for the pions and thus
effects the pion decay constant
\be
f_\pi^2=\frac{M_A^2-M_V^2}{4M_A^2G_2}
\label{fpi}
\ee
with the (axial--) vector meson masses
\be
M_V^2=\frac{g_V^2}{4G_2}\qquad {\rm and}\qquad
M_A^2=M_V^2+6m^2.
\label{vecmass}
\ee
This brief summary of known results has demonstrated the usefulness
of the chiral rotation (\ref{psitilde}) especially in the context
of the derivative expansion since it eliminates the derivative of
the chiral field from the fermion determinant (\ref{rot2}).

The Lagrangian of the hidden gauge approach can be obtained from
(\ref{expand}) by the following approximation. One neglects the
kinetic parts for the axial--vector field $\tilde A_\mu^\prime$ which
leaves this field only as an auxiliary field. This allows to employ
the corresponding equation of motion to eliminate $\tilde A_\mu^\prime$
resulting in
\be
{\cal L}\sim\frac{1}{2g_V^2}{\rm tr}\ \tilde V_{\mu\nu}^2
-a f_\pi^2\ {\rm tr}\ \left(\tilde V_\mu -v_\mu\right)^2
-\frac{1}{4}f_\pi^2\ {\rm tr}\ a_\mu^2.
\label{bando}
\ee
In the work of Bando et al. \cite{ba87} $a$ was left as an undetermined
parameter. Here it is fixed in terms of physical quantities
\be
a=\frac{M_A^2}{M_A^2-M_V^2}.
\label{apara}
\ee
Assuming the constituent quark mass $m=M_V/\sqrt{6}$ not only
yields the Weinberg relation $M_A=\sqrt{2}M_V$ \cite{we67} but
also the KSRF relation $a=2$ \cite{ka66}.

Alternatively one might apply the same manipulations to the
formulation in terms of the unrotated fields (\ref{vecfield}).
Then the chiral field still appears in the fermion determinant
and one has to deal with the covariant derivative
\be
{\cal D}_\mu U = \partial_\mu U +\left[V_\mu,U\right]
-\left\{A_\mu,U\right\}.
\label{covder}
\ee
The leading terms in the Lagrangian can readily be obtained \cite{eb86}
\be
{\cal L}\sim \frac{3m^2}{2g_V^2}{\rm tr}\
\left({\cal D}_\mu U {\cal D}^\mu U^{\dag} \right)
+\frac{1}{2g_V^2}{\rm tr}\
\left(V_{\mu\nu}^2+A_{\mu\nu}^2\right)
-\frac{1}{4G_2}{\rm tr} \left(V_\mu^2+A_\mu^2\right)
\label{expunrot}
\ee
which exactly represent the massive Yang--Mills
Lagrangian \cite{kay84,go84}. Transforming the (axial--) vector
fields according to (\ref{rotvec}) and noting that \cite{wa88}
\be
{\rm tr}\ \tilde A_\mu^2 = \frac{-1}{4}{\rm tr}\
\left({\cal D}_\mu U {\cal D}^\mu U^{\dag} \right)
\label{aeqdu}
\ee
one immediately observes that (\ref{expunrot}) and (\ref{expand})
describe the same physics. In the language of the NJL model the
identity of the HGS and MYM approaches stems from the invariance
of the module of the fermion determinant under the special chiral
rotation (\ref{psitilde}).

After we have seen how the equivalence of the hidden gauge and
massive Yang Mills Lagrangians emerge from the NJL model we
next explore the transformation properties of the fields under
flavor rotations $g_L,\ g_R$. These properties ploy a key role
in order to demonstrate that the elimination of the axial vector
meson field by setting $\tilde A_\mu=0$ does not violate chiral
symmetry. The flavor rotations are defined for the unrotated left--
and right--handed quark fields
\be
\Psi_L\rightarrow g_L\Psi_L\qquad {\rm and} \qquad
\Psi_R\rightarrow g_R\Psi_R.
\label{qtransf}
\ee
The term which describes the coupling of the quarks to
the scalar and pseudoscalar mesons is left invariant by demanding
\be
\xi_L^{\dag} \Sigma\xi_R\rightarrow
g_L\xi_L^{\dag} \Sigma\xi_R g_R^{\dag}
\label{mtransf}
\ee
which introduces the hidden gauge transformation $h$ \cite{eb86}
\be
\xi_L\rightarrow h^{\dag}  \xi_L g_L^{\dag}
\quad , \qquad
\xi_R\rightarrow h^{\dag}  \xi_R g_R^{\dag}
\qquad {\rm and}\qquad
\Sigma\rightarrow h^{\dag} \Sigma h.
\label{xitransf}
\ee
Obviously the scalar fields transform homogeneously under the
hidden gauge transformation. In this context it is important to
note that $h$ may not be chosen independently but rather
depends on the gauge adopted for the chiral fields. Consider
{\it e.g.} the unitary gauge $\xi_L^{\dag} =\xi_R=\xi$.
This requires the transformation property
\be
\xi\rightarrow g_L\xi h = h^{\dag}  \xi g_R^{\dag} .
\label{ugaugetrans}
\ee
For vector type transformations $g_L=g_R=g_V$ this equation is
obviously solved by $h=g_V^{\dag} $. Contrary, for axial type
transformations $g_L=g_R^{\dag} =g_A$ $h$ is obtained as the
solution to $g_A\xi h = h^{\dag}  \xi g_A$ which depends on
the field configuration $\xi$. Thus even for global flavor
transformations $g_{A,V}$ the hidden gauge transformation $h$ may be
coordinate--dependent for coordinate dependent
$\xi(x)$ \cite{ca69,kay84a}. The unrotated (axial--) vector fields
are required to transform inhomogeneously under the flavor rotations
\be
V_\mu+A_\mu\rightarrow
g_R\left(\partial_\mu + V_\mu+A_\mu\right)g_R^{\dag}
\qquad {\rm and}\qquad
V_\mu-A_\mu\rightarrow
g_L\left(\partial_\mu + V_\mu-A_\mu\right)g_L^{\dag} .
\label{vtrans}
\ee
It is then straightforward to verify that the flavor transformation
of the rotated fields only involves the hidden symmetry transformation
$h$ \cite{eb86}
\be
\tilde \Psi_{L,R}\rightarrow h^{\dag}  \tilde \Psi_{L,R}
\quad , \qquad
\tilde V_\mu\rightarrow
h^{\dag} \left(\partial_\mu +\tilde V_\mu\right)h
\qquad {\rm and}\qquad
\tilde A_\mu\rightarrow h^{\dag}  \tilde A_\mu h.
\label{vrottrans}
\ee
The fact that $\tilde A_\mu$ transforms homogeneously has the
important consequence that (as already mentioned) one can set
$\tilde A_\mu=0$ without breaking chiral symmetry bcause the last
relation in (\ref{vrottrans}) does not induce any inhomogeneity.
Furthermore the vector mesons $\tilde V_\mu$ are not affected by this
choice. Let us next examine the NJL model defined by $\tilde A_\mu=0$
in more detail. The corresponding Dirac operator reads
\be
i\tilde{\Dslash} =
i\gamma_\mu\left(\partial^\mu+\tilde{V}^\mu\right)-\Sigma
\label{da1el}
\ee
and the derivative expansion (\ref{expand}) simplifies to
\be
\frac{1}{2g_V^2}{\rm tr}\ \tilde V_{\mu\nu}^2
-\frac{1}{4G_2}{\rm tr} \left[\left(\tilde
V_\mu-v_\mu\right)^2+a_\mu^2\right] + \ldots\ .
\label{exa1el}
\ee
Identifying $\tilde V_\mu$ with the physical $\rho$--meson field
determines the coupling constant $G_2$
\be
m_\rho^2=\frac{g_V^2}{4G_2}=\frac{2\pi^2}
{G_2\Gamma\left(0,\frac{m^2}{\Lambda^2}\right)}
\label{mrg2}
\ee
for $N_C=3$. One may rewrite the last term in equation (\ref{exa1el})
in terms of the chiral field $U$ (see also eqn. (\ref{aeqdu}))
\be
{\rm tr}\ a_\mu^2=-\frac{1}{4}{\rm tr}\
\partial_\mu U\partial^\mu U
\label{nlinsig}
\ee
which provides an additional relation for $G_2$ in terms of the
pion decay constant
\be
f_\pi^2=\frac{1}{4G_2}.
\label{fpia1el}
\ee
This finally implies $m_\rho^2=8\pi^2f_\pi^2/
\Gamma\left(0,m^2/\Lambda^2\right)$. It should be remarked that
the above derived relations are also obtained when (in a derivative
expansion) $A_\mu$ is set to zero and $V_\mu$ identified with the
$\rho$--meson. This, however, is not surprising since these relations
for the vector mesons stem from the part of the Lagrangian which
does not contain pion fields. In the absence the pion fields
$\tilde V_\mu$ and $V_\mu$ are identical, {\it cf.} eqn (\ref{rotvec}).

The derivative expansion (\ref{exa1el}) also contains the
$\rho\pi\pi$ vertex
\be
\frac{g_{\rho\pi\pi}}{\sqrt{2}}\mbox{\boldmath $\rho$}_\mu\cdot
\left(\mbox{\boldmath $\pi$}\times\mbox{\boldmath $\pi$}\right)
=\frac{1}{2G_2}{\rm tr}\left(\tilde V_\mu v^\mu\right).
\label{grpp}
\ee
Expanding the $RHS$ in terms of the physical fields
$\mbox{\boldmath $\rho$}_\mu$ and $\mbox{\boldmath $\pi$}$ yields
\be
g_{\rho\pi\pi}=\frac{m_\rho}{\sqrt{2}f_\pi}
\label{grpp1}
\ee
which falls short a factor $\sqrt{2}$ of the KSRF relation. This
seems to indicate that some relevant information is lost by
setting $\tilde A_\mu=0$. Whether this is also the case in the
soliton sector will be explored in section 5.

In phenomenological vector meson models frequently a term like
\be
{\rm tr}\left[U\left(V_\mu+A_\mu\right)
U^{\dag} \left(V_\mu-A_\mu\right)\right]
\label{addterm}
\ee
is added \cite{kay84a,sch93} to reproduce the empirical value
for $g_{\rho\pi\pi}$ when setting $\tilde A_\mu=0$. This term is
invariant under global chiral rotations. The absence of
this term in the NJL model causes the improper prediction
for this coupling constant.

\bigskip
\stepcounter{chapter}
\leftline{\large\it 3. The NJL soliton}
\bigskip

In the baryon number one sector the NJL model has the celebrated
feature to possess localized static solutions with finite energy,
{\it i.e.} solitons \cite{re89,re88}. Here we wish to briefly
review this solution for the case of pseudoscalar fields.

For static field configurations it is convenient to introduce
a Dirac Hamiltonian
\be
{\cal H}= \mbox {\boldmath $\alpha \cdot p $} +
\beta\left(P_R\xi\langle \Sigma\rangle\xi
+P_L\xi^{\dag}\langle \Sigma\rangle\xi^{\dag}\right)
\label{stham}
\ee
where we have assumed the unitary gauge ({\it i.e.}
$\xi_L^{\dag}=\xi_R=\xi$). This Hamiltonian enters the Euclidean
Dirac operator via
\be
i\beta\Dslash_E=-\partial_\tau-{\cal H}.
\label{eucdirac}
\ee
For static mesonic background fields the fermion determinant can
conveniently be expressed in terms of the eigenvalues $\epsilon_\mu$
of the Dirac Hamiltonian\footnote{The treatment generalizes to
more complicated field configurations as long as ${\cal H}$ is
Hermitian ({\it cf.} section 5).}
\be
{\cal H}\Psi_\mu=\epsilon_\mu\Psi_\mu.
\label{direig}
\ee
These eigenvalues are functionals of the mesonic background
fields. Depending of the specific boundary condition (which fixes
the quantum reference state) to the fermion fields in the functional
integral (\ref{zferm}), the fermion determinant (\ref{zferm})
contains in general besides a vacuum part ${\cal A}^0$ also
a valence quark part ${\cal A}^{\rm val}$ \cite{re89}
\be
{\cal A}={\cal A}^0+{\cal A}^{\rm val}.
\label{actionsum}
\ee
The valence quark part arising from the explicit occupation of
the valence quark levels is given by
\be
{\cal A}^{\rm val}=-E^{\rm val}[\xi] T\ , \quad
E^{\rm val}[\xi]=N_C\sum_\mu \eta_\mu |\epsilon_\mu|.
\label{eval}
\ee
Here $\eta_\mu=0,1$ denote the occupation numbers of the valence
(anti-) quark states. These have to be adjusted such the total baryon
number
\be
B=\sum_\mu \left(\eta_\mu
-\frac{1}{2}{\rm sgn}\left(\epsilon_\mu\right)\right)
\label{bno}
\ee
equals unity. The vacuum part is conveniently evaluated
for infinite Euclidean times $(T\rightarrow\infty)$ which fixes
the vacuum state as the quantum reference state (no valence quark
orbit occupied). For the present considerations it will be sufficient
to evaluate the real vacuum part
\be
{\cal A}^0_R=\frac{1}{2} {\rm Tr}\ {\rm log} \Dslash^{\dag} _E
\Dslash_E.
\label{areal}
\ee
Since for static configurations one has $[\partial_\tau,{\cal H}]$=0
and thus $\Dslash_E ^{\dag}\Dslash_E = -\partial_\tau^2 +{\cal H}^2$.
Then it is straightforward to evaluate the real part of the
fermion determinant in proper--time regularization\footnote{The
imaginary part does not contribute for the field configurations under
consideration because ${\cal H}$ is assumed to be Hermitian.} \cite{re89}
\be
\A^0_R &=& -\frac{1}{2}\int_{1/\Lambda ^2}^\infty \frac {ds}{s}
{\rm Tr}\ {\rm exp}
\left( -s \Dslash_E ^{\dag}\Dslash_E \right)
\nonumber \\
&=& -T \frac{N_C}{2}\int_{-\infty}^\infty \frac{dz}{2\pi}\sum_\mu
\int_{1/\Lambda ^2}^\infty \frac {ds}{s}
{\rm exp}\left(-s\left[z^2+\epsilon_\mu^2\right]\right).
\label{star}
\ee
The temporal part of the trace has become the $z$ integration. As
this integral is Gaussian it can readily be carried out yielding
\be
\A^0_R=-T \frac{N_C}{2}\int_{1/\Lambda ^2}^\infty
\frac{ds}{\sqrt{4\pi s^3}}\sum_\mu
{\rm exp}\left(-s\epsilon_\mu^2\right).
\label{arsum}
\ee
This expression allows to read off the static energy functional $E[\xi]$
since $\A^0_R\rightarrow -T E^0[\xi]$ as $T\rightarrow\infty$
\be
E^0[\xi]=\frac{N_C}{2}\int_{1/\Lambda ^2}^\infty
\frac{ds}{\sqrt{4\pi s^3}}\sum_\mu
{\rm exp}\left(-s\epsilon_\mu^2\right).
\label{evac}
\ee
Finally the total energy functional is given by
\be
E[\xi]=E^{\rm val}[\xi]+E^0[\xi]-E^0[\xi=1]
\label{etot}
\ee
which is normalized to the energy of the vacuum configuration
$\xi=1$. In the chiral limit ($m_\pi=0$), which we have
adopted here, the meson part of the action does not contribute
to the soliton energy. The chiral soliton is the $\xi$ configuration
which minimizes $E[\xi]$ and the minimal $E[\xi]$ is then identified
as the soliton mass.

To be specific we employ the hedgehog {\it ansatz} for the chiral
field
\be
\xi(\mbox{\boldmath $r$})={\rm exp}\left(\frac{i}{2}
{\mbox{\boldmath $\tau$}} \cdot{\hat{\mbox{\boldmath $r$}}}\
\Theta(r)\right)
\label{chsol}
\ee
while the scalar fields are constrained to the chiral circle,
{\it i.e.} $\langle \Sigma\rangle =m$. Substituting this
{\it ansatz} into the Dirac Hamiltonian (\ref{stham}) yields
\be
{\cal H}&=&
{\mbox {\boldmath $\alpha$}} \cdot {\mbox{\boldmath $p$}} +
\beta m \left({\rm cos}\Theta(r) + i\gamma_5{\mbox{\boldmath $\tau$}}
\cdot{\hat{\mbox{\boldmath $r$}}}\ {\rm sin}\Theta(r)\right).
\label{h0}
\ee

The stationary condition $\delta E[\xi]/\delta\xi=0$ is made
explicit by functionally differentiating the energy--eigenvalues
$\epsilon_\mu$ with respect to $\Theta$
\be
\frac{\delta\epsilon_\mu}{\delta\Theta(r)}=
m \int d\Omega\  \Psi_\mu^{\dag} (\mbox{\boldmath $r$})\beta
\left(-{\rm sin}\Theta(r)+i\gamma_5{\mbox{\boldmath $\tau$}}
\cdot{\hat{\mbox{\boldmath $r$}}}\ {\rm cos}\Theta(r)\right)
\Psi_\mu(\mbox{\boldmath $r$}).
\label{diffeps}
\ee
This leads to the equation of motion \cite{re88}
\be
{\rm cos}\Theta(r)\  {\rm tr}\int d\Omega\
\rho_S(\mbox{\boldmath $r$},\mbox{\boldmath $r$})
i\gamma_5\mbox{\boldmath $\tau$} \cdot{\hat{\mbox{\boldmath $r$}}}\ =
{\rm sin}\Theta(r)\  {\rm tr}\int d\Omega\
\rho_S(\mbox{\boldmath $r$},\mbox{\boldmath $r$})
\label{eqm}
\ee
where the traces are over flavor and Dirac indices only. According to
the sum (\ref{etot}) the scalar quark density matrix
$\rho_S(\mbox{\boldmath $x$},\mbox{\boldmath $y$}))=
\langle q(\mbox{\boldmath $x$})\bar q(\mbox{\boldmath $y$})\rangle$
is decomposed into valence quark and Dirac sea parts:
\be
\rho_S(\mbox{\boldmath $x$}, \mbox{\boldmath $y$}) & = &
\rho_S^{\rm val}(\mbox{\boldmath $x$},\mbox{\boldmath $y$})
+ \rho_S^{\rm vac}(\mbox{\boldmath $x$},\mbox{\boldmath $y$})
\nonumber \\*
\rho_S^{\rm val}(\mbox{\boldmath $x$},\mbox{\boldmath $y$}) & = &
\sum _\mu
\Psi_\mu(\mbox{\boldmath $x$})\eta_\mu
\bar \Psi_\mu(\mbox{\boldmath $y$}) {\rm sgn} (\epsilon_\mu)
\nonumber \\*
\rho_S^{\rm vac}(\mbox{\boldmath $x$},\mbox{\boldmath $y$}) & = &
\frac{-1}{2}\sum_\mu \Psi_\mu(\mbox{\boldmath $x$})
{\rm erfc}\left(\left|\frac{\epsilon_\mu}{\Lambda}\right|\right)
\bar \Psi_\mu(\mbox{\boldmath $y$}) {\rm sgn} (\epsilon_\mu) .
\label{density}
\ee
The explicit form of the eigenfunctions $\Psi_\mu(\mbox{\boldmath
$x$})$ as well as remarks on the appropriate boundary conditions
may be found in the appendix.

\bigskip
\stepcounter{chapter}
\leftline{\large\it 4.The chirally rotated fermion determinant}
\bigskip

In the appendix it is demonstrated that the normalizable
solutions to the free Dirac equation with spherical boundary conditions
represent a pertinent basis for the diagonalization of the Dirac
Hamiltonian with the soliton present. This property is based on the
fact that the Hamiltonian (\ref{h0}) is free of singularities. In this
section we will explain how singularities appearing in a Dirac
Hamiltonian influence the choice of basis states. Let us for this
purpose consider the Hamiltonian for the chirally rotated
quark fields $\tilde \Psi=\Omega(\Theta)\Psi$
({\it cf.} eqns (\ref{wuyang})--(\ref{atheta}) and ref. \cite{re89a}):
\be
{\cal H}_R= \Omega(\Theta){\cal H}\Omega^{\dag} (\Theta)&=&
\mbox{\boldmath $\alpha$}\cdot\mbox{\boldmath $p$}
+\beta m -\frac{1}{2}(\mbox{\boldmath $\sigma$}\cdot
\hat{\mbox{\boldmath $r$}})
(\mbox{\boldmath $\tau$}\cdot\hat{\mbox{\boldmath $r$}})
\left(\Theta^\prime(r)-\frac{1}{r}\sin\Theta(r)\right) \nonumber \\
 & &-\frac{1}{2r}(\mbox{\boldmath $\sigma$}\cdot\mbox{\boldmath $\tau$})
\sin\Theta(r)-\frac{1}{r}\mbox{\boldmath $\alpha$}\cdot
(\hat{\mbox{\boldmath $r$}}\times\mbox{\boldmath $\tau$})
\sin^2\left(\frac{\Theta(r)}{2}\right)
\label{hrot}
\ee
since in unitary gauge
$\Omega(\Theta)={\rm cos}(\Theta/2)
+i\gamma_5\mbox{\boldmath $\tau$}\cdot\hat{\mbox{\boldmath $r$}}\
{\rm sin}(\Theta/2)$.
Obviously the $\Theta$--dependence in the Hamiltonian has been
transferred to induced (axial--) vector meson fields. As expected the
rotated Hamiltonian, ${\cal H}_R$, contains an explicit singularity in
the $1/r\mbox{\boldmath $\alpha$}\cdot
(\hat{\mbox{\boldmath $r$}}\times\mbox{\boldmath $\tau$})
\sin^2\left(\frac{\Theta(r)}{2}\right)$ term at $r=0$.
Additionally there are ``coordinate singularities" in the
expressions involving $\hat{\mbox{\boldmath $r$}}$. All these
singularities appear because the ``coordinate singularity" in
$\Omega(\Theta)$ is not compensated by corresponding values of the
chiral angle $\Theta$. Stated otherwise: the transformation
$\Omega(\Theta)$ with $\Theta(0)-\Theta(\infty)=-n\pi$ is
topologically distinct from the unit transformation. Although
$\Omega(\Theta)$ represents a unitary transformation it is then
not astonishing that a numerical diagonalization
\be
{\cal H}_R \tilde \Psi_\mu =\tilde\epsilon_\mu \Psi_\mu
\label{dhrot}
\ee
in the basis of the free Hamiltonian does ${\underline {\rm not}}$ render
the eigenvalues of the unrotated Hamiltonian, ${\cal H}$, ({\it i.e.}
$\tilde\epsilon_\mu\ne \epsilon_\mu)$ despite the relevant
matrix elements being finite.
This finiteness is merely due to the $r^2$ factor in the volume element.
One might suspect that the Hamiltonian ${\cal H}_{2R}=
\Omega(2\Theta){\cal H} \Omega^{\dag} (2\Theta)$ obtained by a
$2\Theta$ rotation has the same spectrum as ${\cal H}$, since
${\cal H}_{2R}$ is free of singularities. Although this behavior is
exhibited by the numerical solution for the low--lying energy
eigenvalues, the topological character of the transformation has
drastic consequences for the states at the lower and upper ends of the
spectrum in momentum space. Adopting the same basis states for
diagonalizing ${\cal H}$ and ${\cal H}_{2R}$  one observes that the
eigenvalues of ${\cal H}_{2R}$ are shifted against those of ${\cal H}$,
{\it i.e.} the most negative energy eigenvalue is missing while an
additional one has popped up at the upper end of the spectrum. Up
to numerical uncertainties the eigenvalues in the intermediate region
agree for both ${\cal H}$ and ${\cal H}_{2R}$. This behavior is
sketched in figure 4.1 and repeats itself for
${\cal H}_{4R}=\Omega(4\Theta){\cal H}\Omega^{\dag} (4\Theta)$.
Thus the chiral rotation represents another example of the
so--called ``infinite hotel story" \cite{ni83} which is an
interesting feature reflecting the topological character of this
transformation.

Let us now return to the problem of diagonalizing ${\cal H}_R$ and
restrict ourselves for the moment to the channel $G^\Pi=0^+$.
The grand spin operator $\mbox{\boldmath $G$}$ is defined in the
appendix (\ref{gspin}). It should be stressed that
$[\Omega,\mbox{\boldmath $G$}]=0$. Thus the local chiral rotation
may be investigated in each grand spin channel separately.

At $r=0$ the chiral rotation
\be
\Omega(r=0)=-i\left(\mbox{\boldmath $\tau$}\cdot
\hat{\mbox{\boldmath $r$}}\right)\gamma_5
\label{globalrot}
\ee
obviously exchanges upper and lower components of Dirac
spinors\footnote{In the standard representation
$\gamma_5=\pmatrix{0 & 1\cr 1 &0\cr}$.}. The corresponding
wave--functions $\tilde\Psi_\mu^{(0,+)}(r=0)=
\Omega(r=0)\Psi_\mu^{(0,+)}(r=0)$ (see eqn (\ref{psi0rot})) can
obviously not be represented by the free basis as in general the
lower component of $tilde\Psi_\mu^{(0,+)}(r=0)$ is different from
zero. However, the lower components of the eigenstates of the free
Hamiltonian in the $G^\Pi=0^+$ channel have this property ({\it cf.}
eqns (\ref{psipos},\ref{freesol})). It is thus clear that the local
chiral rotation is not unitary in a finite model space. Furthermore
the equality  $\epsilon_\mu=\tilde\epsilon_\mu$ cannot be gained
without additional manipulations. Of course, this ``non--unitarity"
is completely due to the topological character of this rotation.
This problem can be avoided by defining a basis in the
topologically non--trivial sector via
\be
\tilde \Psi_{\mu 0}=\Omega (\phi)\Psi_{\mu 0}
\label{topbasis}
\ee
with $\Psi_{\mu 0}$ being the solutions to the free unrotated
Hamiltonian. $\phi$ represents an auxiliary  radial function satisfying
the boundary conditions $\phi(0)=-\pi$ and $\phi(D)=0$. {\it E.g.}
we may take\footnote{$D$ denotes the size of the spherical cavity which
serves to discretize the momentum eigenstates, {\it cf.} the appendix.}
\be
\phi(r)=-\pi\left(1-\frac{r}{D}\right)
{\rm exp}\left(-tmr\right)
\label{ansatzphi}
\ee
with $t$ being a free parameter. The diagonalization of ${\cal H}_R$
in the basis $\tilde \Psi_0$ is equivalent to diagonalizing
\be
&&\hspace{2cm}
\mbox{\boldmath $\alpha$}\cdot\mbox{\boldmath $p$}
+m \beta \left(\cos\phi(r)+i\gamma_5
\mbox{\boldmath $\tau$}\cdot\hat{\mbox{\boldmath $r$}}
\sin\phi(r)\right) \nonumber \\
&&\hspace{1cm}
+\frac{1}{2}\left[\phi^\prime(r)-\Theta^\prime(r)
+\frac{1}{r}\sin(\Theta(r)-\phi(r))\right]
\mbox{\boldmath $\sigma$}\cdot\hat{\mbox{\boldmath $r$}}
\mbox{\boldmath $\tau$}\cdot\hat{\mbox{\boldmath $r$}}
\nonumber \\
&&+\frac{1}{2r}\sin(\phi(r)-\Theta(r))
\mbox{\boldmath $\sigma$}\cdot\mbox{\boldmath $\tau$}
-\frac{1}{2r}\left[1-\cos(\Theta(r)-\phi(r))\right]
\mbox{\boldmath $\alpha$}\cdot
(\hat{\mbox{\boldmath $r$}}\times\mbox{\boldmath $\tau$})
\label{hphi}
\ee
in the standard basis $\{\Psi_{\mu0}\}$. At this point
it should be stressed again that $\phi$ is not a dynamical field
but merely an auxiliary field which transforms the basis such as
to eliminate the $1/r$--singularities from the Dirac Hamiltonian.
This property is completely determined by the boundary values of
$\phi$. Thus the results ought to be independent of the parameter
$t$. We have confirmed this property numerically. The total energy
of the soliton varies only by fractions of MeV in a wide range of $t$.
This is, of course, negligibly small since the inherited mass scale is
given by the constituent quark mass, $m$ which is of the order of several
hundred MeV. By using the locally transformed basis (\ref{topbasis})
also the wave--functions corresponding to the eigenstates of ${\cal H}_R$
agree reasonably well with the rotated wave--functions $\Omega(\Theta)
\Psi_\mu$ of the original Dirac Hamiltonian ${\cal H}$. It should be
noted that a large number of momentum states is required to numerically
gain this result. This is not surprising since in order to represent
the functional unity an infinite number of momentum states is needed.

At this point we want to add a remark on a further application
of the chiral rotation. Using $\Theta\equiv\pm\pi$,
{\it i.e.} a chiral transformation $\Omega=\pm i\gamma_5
\mbox{\boldmath $\tau$}\cdot\hat{\mbox{\boldmath $r$}}$, the
rotated Hamiltonian (\ref{hrot}) is given by
\be
{\cal H}_R^{\Theta=\pm\pi}=\mbox{\boldmath $\alpha$}\cdot
\mbox{\boldmath $p$}+\beta m-\frac{1}{2}\mbox{\boldmath $\alpha$}
\cdot\left(\hat{\mbox{\boldmath $r$}}\times
\mbox{\boldmath $\tau$}\right)
\label{hrtconst}
\ee
whereas the unrotated one (see eqn (\ref{h0})) is simply
${\cal H}^{\Theta=\pm\pi}=\mbox{\boldmath $\alpha$}\cdot
\mbox{\boldmath $p$}-\beta m$. The latter is diagonalized
straightforwardly. It has the same eigenvalues as the free
Hamiltonian and the eigenfunctions $\Psi^{\Theta=\pm\pi}$
differ by the substitution $m\rightarrow-m$. Therefore the
Hamiltonian (\ref{hrtconst}) has also the eigenvalues of the
free Hamiltonian. The eigenfunctions are given by
$\Psi_R^{\Theta=\pm\pi}= \Omega^{\dag} (\Theta=\pm\pi)
\Psi^{\Theta=\pm\pi}=\pm i\gamma_5\mbox{\boldmath $\tau$}\cdot
\hat{\mbox{\boldmath $r$}}\Psi^{\Theta=\pm\pi}$. {\it I.e.}
the eigenfunctions of (\ref{hrtconst}) are easily constructed:
start with the free eigenfunctions (\ref{psipos}-\ref{freesol}),
substitute $m\rightarrow-m$ and apply $\pm i\gamma_5
\mbox{\boldmath $\tau$}\cdot\hat{\mbox{\boldmath $r$}}$. This
constitutes just one example how the chiral rotation (\ref{psitilde})
can be used to diagonalize non--trivial operators.

Before turning to the detailed discussion of the equation of motion in
the rotated system we would like to mention that the self--consistent
profile being obtained from the unrotated problem also minimizes the
soliton mass in the chirally rotated frame. Stated otherwise: each
change in this profile function leads to an increase of the energy
obtained from the eigenvalues of ${\cal H}_R$.

As in the formulation within the unrotated frame the equation
of motion is gained by extremizing the energy functional (\ref{etot}).
The energy eigenvalues in the rotated frame, however, exhibit
a different functional dependence on the chiral field. The functional
derivative of these eigenvalues with respect to the chiral angle
reads
\be
\frac{\delta\epsilon_\mu}{\delta\Theta(r)}&=&
\int d\Omega\ \Big\{\frac{1}{2}\Big(\frac{\partial}{\partial r}
+\frac{1}{r}{\rm cos}\Theta(r)\Big) r^2 \tilde\Psi_\mu^{\dag}
(\mbox{\boldmath $r$}) \mbox{\boldmath $\sigma$}\cdot
\hat{\mbox{\boldmath $r$}}
\mbox{\boldmath $\tau$}\cdot\hat{\mbox{\boldmath $r$}}
\tilde\Psi_\mu(\mbox{\boldmath $r$})
\label{diffrot}
\\ &&  \qquad
-\frac{r}{2}{\rm cos}\Theta(r)\tilde\Psi_\mu^{\dag}
(\mbox{\boldmath $r$})
\mbox{\boldmath $\sigma$}\cdot\mbox{\boldmath $\tau$}
\tilde\Psi_\mu(\mbox{\boldmath $r$})
-\frac{r}{2}{\rm sin}\Theta(r)\tilde\Psi_\mu^{\dag}
(\mbox{\boldmath $r$})
\mbox{\boldmath $\alpha$}\cdot
(\hat{\mbox{\boldmath $r$}}\times\mbox{\boldmath $\tau$})
\tilde\Psi_\mu(\mbox{\boldmath $r$})\Big\}
\nonumber
\ee
with $\tilde\Psi_\mu(\mbox{\boldmath $r$})$ being the eigenstates
of the rotated Dirac Hamiltonian (\ref{hrot}). The derivatives
(\ref{diffrot}) enter the stationary condition for the energy
functional resulting in the equation of motion
\be
A_L(r)+A_T(r){\rm cos}\Theta(r)-V(r){\rm sin}\Theta(r)=0.
\label{eqmrot}
\ee
In order to display the radial functions $A_L,A_T$ and $V$ it
is convenient to introduce the charge density $\tilde\rho_C=
\langle q(\mbox{\boldmath $x$}) q(\mbox{\boldmath $y$})^{\dag}
\rangle=\tilde\rho_C^{\rm val}+\tilde\rho_C^{\rm vac}$ involving the
eigenstates $\tilde\Psi_\mu$ of ${\cal H}_R$ ({\it cf.} eqn
(\ref{density})) \cite{re89}:
\be
\tilde\rho_C^{\rm val}(\mbox{\boldmath $x$},\mbox{\boldmath $y$})
& = & \sum _\mu \tilde\Psi_\mu(\mbox{\boldmath $x$})\eta_\mu
\tilde\Psi_\mu^{\dag} (\mbox{\boldmath $y$})
{\rm sgn} (\epsilon_\mu)
\nonumber \\*
\tilde\rho_C^{\rm vac}(\mbox{\boldmath $x$},\mbox{\boldmath $y$})
& = & \frac{-1}{2}\sum_\mu \tilde\Psi_\mu(\mbox{\boldmath $x$})
{\rm erfc}\left(\left|\frac{\epsilon_\mu}{\Lambda}\right|\right)
\tilde\Psi_\mu^{\dag} (\mbox{\boldmath $y$})
{\rm sgn} (\epsilon_\mu) .
\label{densrot}
\ee
The radial functions $A_L,A_T$ and $V$ are of axial--($A_{L,T}$)
and vector($V$) character
\be
A_L(r)&=&\frac{1}{r}\frac{\partial}{\partial r}\
{\rm tr}\ \int d\Omega\ r^2\ \tilde\rho_C
(\mbox{\boldmath $r$},\mbox{\boldmath $r$})
\mbox{\boldmath $\sigma$}\cdot\hat{\mbox{\boldmath $r$}}
\mbox{\boldmath $\tau$}\cdot\hat{\mbox{\boldmath $r$}}
\label{defal} \\
A_T(r)&=&{\rm tr}\ \int d\Omega\ \tilde\rho_C
(\mbox{\boldmath $r$},\mbox{\boldmath $r$})
\left[\mbox{\boldmath $\sigma$}\cdot\hat{\mbox{\boldmath $r$}}
\mbox{\boldmath $\tau$}\cdot\hat{\mbox{\boldmath $r$}}-
\mbox{\boldmath $\sigma$}\cdot\mbox{\boldmath $\tau$}\right]
\label{defat} \\
V(r)&=&{\rm tr}\ \int d\Omega\ \tilde\rho_C
(\mbox{\boldmath $r$},\mbox{\boldmath $r$})
\mbox{\boldmath $\alpha$}\cdot
(\hat{\mbox{\boldmath $r$}}\times\mbox{\boldmath $\tau$}).
\label{defv}
\ee
It should be noted that $A_L$ does not depend on the auxiliary field
$\phi$. It is then straightforward to verify that for any meson
configuration $A_L$ and $A_T$ satisfy the relation
\be
A_L(r=0)=-(-1)^kA_T(r=0)
\label{aorigin}
\ee
where $k$ is defined by the value of the auxiliary field $\phi$ at
the origin $\phi(r=0)=k\pi$. The vector type radial function
$V$ vanishes at the origin. Thus the equation of motion
(\ref{eqmrot}) together with the relation (\ref{aorigin}) yield
the boundary condition $\Theta(r=0)=(2n+1)\pi$ for $k=1$. This is
stronger than the boundary condition derived from the original equation
of motion (\ref{eqm}) which also allows for even multiples of $\pi$
for $\Theta(r=0)$ since $\tr \int d\Omega\ \rho_S
(\mbox{\boldmath $r$}=0,\mbox{\boldmath $r$}=0)
i\gamma_5 \mbox{\boldmath $\tau$}\cdot\hat{\mbox{\boldmath $r$}}=0$
Assuming the Kahana--Ripka \cite{ka84} boundary conditions for the
unrotated basis states $\Psi_{\mu0}$ similar considerations for $r=D$
show that $\Theta(r=D)=2l\pi$ since $A_L(r=D)=-A_T(r=D)$ as long as
$\phi(D)=0$. Obviously the topological charge associated with the
chiral field in the hedgehog {\it ansatz} $(\Theta(r=0)-\Theta(r=D))/\pi$
can assume odd values only when $k$ is odd. In particular $\phi\equiv0$
is prohibited in the case of unit baryon number. Thus the study of the
boundary conditions in the chirally transformed system corroborates the
conclusion drawn from investigating the eigenvalues and --states
of ${\cal H}_R$ that it is mandatory to also transform the basis
spinors and in particular the boundary conditions to the topological
non--trivial sector.

We would also like to remark that this kind of singularities does
not only appear when the Dirac Hamiltonian is considered. Such
topological defects have also caused problems in Skyrme type models when
fluctuations off vector meson solitons were investigated \cite{sch89}.
In that case the boundary conditions for the vector meson fluctuations
had to undergo a special gauge transformation which corresponds to the
transformation of the basis quark spinors described here (\ref{topbasis}).

Before discussing the numerical treatment of eqn (\ref{eqmrot}) we
would like to make the remark that substituting the transformation
$\Psi_\mu(\mbox{\boldmath $r$})=\Omega^{\dag}  (\Theta)
\tilde\Psi_\mu(\mbox{\boldmath $r$})$ into the original equation of
motion (\ref{eqm}) does not result in the relation (\ref{diffrot})
but rather yields the constraint
\be
0=\sum_\mu\left(\eta_\mu {\rm sgn}(\epsilon_\mu)
-\frac{1}{2}{\rm erfc}\left(\left|
\frac{\epsilon_\mu}{\Lambda}\right|\right)\right)
\bar{\tilde\Psi}_\mu(\mbox{\boldmath $r$}) \gamma_5
\mbox{\boldmath $\tau$}\cdot\hat{\mbox{\boldmath $r$}}
\tilde\Psi_\mu(\mbox{\boldmath $r$}).
\label{constr}
\ee
Thus the equation of motion (\ref{eqmrot}) cannot be obtained by
transforming the states $\Psi_\mu$ but only by employing the
Dirac equation in the rotated frame (\ref{dhrot}) to extremize the
energy functional. The constraint (\ref{constr}) does not represent
an over--determination of the system since infinitely many
states are involved.

Due to the transcendent character of the equation of motion in the rotated
frame (\ref{eqmrot}) solutions cannot be obtained for arbitrary values
of the radial functions $A_L,A_T$ and $V$. {\it E.g.} for large
distances $\Theta\rightarrow0$ requires $|A_L|\le|A_T|$ in order
to find a solution to eqn (\ref{eqmrot}). Thus the treatment of the
NJL soliton in the chirally rotated frame is not very well suited for
an iterative procedure to find the self--consistent solution. The reason
is that a small deviation of the radial functions $A_L,A_T$ and $V$ from
those corresponding to this solution can render eqn (\ref{eqmrot})
indissoluble for $\Theta(r)$. Then it is not unexpected that at
large distances the solution to the rotated equation of motion
(\ref{eqmrot}) becomes unstable and the original profile function
cannot be reproduced for $r\ge2$fm. For smaller values of $r$ the
original profile function is well reproduced. In figure 4.2 this behavior
is displayed. The self--consistent solution to eqn (\ref{eqm})
serves as ingredient to evaluate the radial functions $A_L,A_T$ and $V$.
The solution to eqn (\ref{eqmrot}) is then constructed and compared
to the original profile function.

\bigskip
\stepcounter{chapter}
\leftline{\large\it 5. The soliton without axial vector mesons}
\bigskip

In this section we will examine the soliton solutions when the chirally
rotated axial vector field is set to zero, $\tilde A_\mu=0$. In Skyrme
type models this approach has been frequently used to study the
structure of baryons \cite{ja88,me88,sch89}. Here will ignore
the effects of the $\omega$ meson. There are two reasons for doing
so. First, as this field represents an isosinglet it is not affected
by the chiral rotation. The second reason is of more technical nature.
Since the corresponding grand spin zero ansatz introduces a
non--vanishing temporal component of a vector field the Dirac
Hamiltonian in Euclidean space is no longer Hermitian. A stringent
derivation of the corresponding Minkowski energy functional has
unfortunately not yet been performed successfully. Several attempts
for motivating such an energy functional have, however, been made
\cite{wa92,sch93b,zu94a,we94}. Excluding the $\omega$--meson we have
besides the hedgehog ansatz for the chiral angle (\ref{hedgehog}) only
the Wu--Yang form for the chirally rotated vector field
\be
\tilde V_\mu=(i\omega_\mu,i\tilde v_m^a\frac{\tau^a}{2})_\mu
\label{notcomp}
\ee
wherein
\be
\omega_\mu=0\quad {\rm and}\quad
\tilde v_m^a=\frac{G(r)}{r}\epsilon_{mka}\hat{r}_k.
\label{wuyg}
\ee
Here $G(r)$ refers to the dynamical $\rho$--meson field and should
not be confused with the induced vector field discussed in
eqn (\ref{gtheta}). From the special form of the Dirac operator in
the chirally rotated frame (\ref{da1el}) it is then obvious that
the chiral field only appears in the purely mesonic part
of the energy functional
\be
E_{\rm m}=\frac{\pi}{G_2}\int dr \left\{
\left(G(r)+1-{\rm cos}\Theta(r)\right)^2+\frac{1}{2}r^2
\left(\Theta^\prime(r)\right)^2+{\rm sin}^2\Theta(r)\right\}.
\label{emvec}
\ee
The boundary value $\Theta(r=0)=-\pi$ then implies that
$G(r=0)=-2$ \cite{ja88,me88}.  The stationary condition for the
chiral angle now becomes a second order non--linear differential equation
\be
\frac{d^2\Theta(r)}{dr^2}=-\frac{2}{r}\frac{d\Theta(r)}{dr}
+\frac{1}{r^2}{\rm sin}2\Theta(r)+
\frac{2}{r^2}{\rm sin}\Theta(r)
\left(G(r)+1-{\rm cos}\Theta(r)\right).
\label{diffeqn}
\ee
Once a profile function $G(r)$ is provided this equation can
be solved analogously to Skyrme model calculations.

Undoing the chiral rotation after having set $\tilde A_\mu=0$
yields a Dirac Hamiltonian which contains all possible
grand spin zero operators of positive parity
\be
{\cal H}&=&\mbox{\boldmath $\alpha$}\cdot\mbox{\boldmath $p$}
+m\beta\left({\rm cos}\Theta+i\gamma_5
\hat{\mbox{\boldmath $r$}}\cdot\mbox{\boldmath $\tau$}
{\rm sin}\Theta\right)
+\frac{1}{2}\Theta^\prime
\hat{\mbox{\boldmath $r$}}\cdot\mbox{\boldmath $\sigma$}
\hat{\mbox{\boldmath $r$}}\cdot\mbox{\boldmath $\tau$}
\label{hthrho} \\
&& +\frac{{\rm sin}\Theta}{2r}\left(G+1\right)
\left({\mbox{\boldmath $\tau$}}\cdot\mbox{\boldmath $\sigma$}
-\hat{\mbox{\boldmath $r$}}\cdot\mbox{\boldmath $\sigma$}
\hat{\mbox{\boldmath $r$}}\cdot\mbox{\boldmath $\tau$}\right)
+\frac{1}{2r}\left(G{\rm cos}\Theta-1+{\rm cos}\Theta\right)
\mbox{\boldmath $\alpha$}\cdot
(\hat{\mbox{\boldmath $r$}}\times\mbox{\boldmath $\tau$}).
\nonumber
\ee
In account of the above mentioned boundary values for $\Theta$ and
$G$ this operator does not contain any coordinate singularities.
On the contrary, the chirally rotated Dirac Hamiltonian becomes as
simple as
\be
{\cal H}_R=\mbox{\boldmath $\alpha$}\cdot\mbox{\boldmath $p$}
+\beta m +\frac{G(r)}{2r}\mbox{\boldmath $\alpha$}\cdot
(\hat{\mbox{\boldmath $r$}}\times\mbox{\boldmath $\tau$})
\label{hrvec}
\ee
being, however, singular at $r=0$. Most importantly eqn (\ref{hrvec})
demonstrates that the eigenvalues of ${\cal H}$ (or ${\cal H}_R$),
$\epsilon_\mu$, which enter the energy functional, are indeed
independent of the chiral angle $\Theta$. Then also the contribution
of the fermion determinant to the energy functional $E^{\rm det}=
E^0+E^{\rm val}$ does not dependent on the chiral angle $\Theta$,
{\it i.e} $E^{\rm det} [\Theta,G]=E^{\rm det}[G]$. Here $E^{\rm det}$
is understood as the sum of the eigenvalues in the form
displayed in eqns (\ref{eval},\ref{evac}). In particular an
infinitesimal change of $\Theta$ leaves $E^{\rm det}$ unchanged. Thus
the fermion determinant does not contribute to the equation of motion
for $\Theta$ (\ref{diffeqn}). In view of the singular character
of ${\cal H}_R$ this operator cannot be treated using the standard
basis \cite{ka84,we92} but rather by employing techniques which
are analogous to those developed in section 4. This corresponds to
use the form (\ref{hthrho}) with $\Theta$ substituted by a
reference profile $\phi$ in order to compute $E^{\rm det}[G]$.
Fortunately, as the only relevant information concerns the boundary
values of $\phi$ rather than its explicit form, we may equally well
employ (\ref{hthrho}) with $\Theta$ being the solution of
(\ref{diffeqn}).

As the fermion determinant is a functional of $G$ it contributes to
the associated equation of motion
\be
G(r)={\rm cos}\Theta(r)-1-\frac{N_C}{8\pi f_\pi^2}\int d\Omega
\sum_\mu\tilde\Psi_\mu^{\dag} (\mbox{\boldmath $r$})
\frac{\mbox{\boldmath $\alpha$}}{2}\cdot
\left(\mbox{\boldmath $r$}\times\mbox{\boldmath $\tau$}\right)
\tilde\Psi_\mu(\mbox{\boldmath $r$})f^\prime(\epsilon_\mu).
\label{eqng}
\ee
where $\epsilon_\mu$ and $\tilde\Psi_\mu(\mbox{\boldmath $r$})$
are the eigenvalues and --functions of ${\cal H}_R$. Furthermore
\be
f^\prime(\epsilon_\mu)=\eta_\mu{\rm sgn}(\epsilon_\mu)
-\frac{1}{2}{\rm sgn}(\epsilon_\mu){\rm erfc}\left(\left|
\frac{\epsilon_\mu}{\Lambda}\right|\right)
\label{eprime}
\ee
denotes the derivative of the fermion determinant with respect to
$\epsilon_\mu$.

Before discussing the emergence of self--consistent solutions we
will present a computation of the fermion determinant associated
with the Dirac Hamiltonian ${\cal H}_R$ for a given profile function
$G(r)$. A physically motivated profile is given by the local
approximation\footnote{We adopt this notation from the analogous
approximation in the Skyrme model which relates the Skyrme term
to the field strength tensor of the $\rho$--meson.} (l.a.) to the
equation of motion (\ref{eqng}) (see also eqn (\ref{gtheta}))
\be
G^{\rm l.a.}(r)={\rm cos}\Theta_{\rm s.c.}(r)-1
\label{locapp}
\ee
wherein $\Theta_{\rm s.c.}(r)$ refers to the self--consistent solution
to the model with pseudoscalar fields only, which is described in
section 3. In the local approximation the mesonic part of the energy
is given by
\be
E^{\rm l.a.}_{\rm m}=
\frac{\pi}{2G_2}\int dr \left\{
r^2\left(\Theta_{\rm s.c.}^\prime(r)\right)^2
+2{\rm sin}^2\Theta_{\rm s.c.}(r)\right\}.
\label{emla}
\ee
The energy eigenvalues are obtained by diagonalizing
${\cal H}$ with $\Theta$ being substituted with the auxiliary
field $\phi$ as given in eqn (\ref{ansatzphi}). For the numbers
listed in table 5.1 we have confirmed that these are independent
of the parameter $t$.

\begin{table}
{}~
\newline
\tcaption{Contributions to the energy for the local approximation
(\ref{locapp}) as functions of the constituent quark mass $m$. All
numbers are in MeV.}
{}~
\newline
\centerline{
\begin{tabular}{c|c c c}
$m$ & ~~400~~ & ~~500~~ & ~~600~~ \\
\hline
$\epsilon_{\rm val}$ & -399  & -500 & -599 \\
$E^{\rm det}[G^{\rm l.a.}]$ & 150  & 84 & 61 \\
$E_m^{\rm l.a.}$ & 1221 & 1513 &  1480 \\
$E_{\rm tot}^{\rm l.a.}$ & 1372 & 1597 & 1541 \\
\end{tabular}}
\end{table}

Obviously the valence quark orbit is extremely bound, its energy
eigenvalue being approximately $-m$\footnote{Here it should be noted
that, as a consequence of discretizing the momentum space, the
smallest module of an energy eigenvalue for a free quark is
$\sqrt{M^2+(\pi/D)^2}$. The eigenvalues of free quarks are
discussed in the appendix.}. Numerically we observe that the upper
component of the corresponding wave--function vanishes. Thus the
valence quark orbit carries all features of an antiquark state.
Nevertheless, the polarization of the Dirac sea only gives a minor
contribution to the energy. This can be understood by noticing that
a shift of the valence quark level from $\epsilon_{\rm val}\approx m$
to $\epsilon_{\rm val}\approx-m$ does not change the vacuum part
of the energu where all energy levels enter with their module, see
eqn (\ref{evac}). Some caution has to be taken when considering the
local approximation. It may be far off the actual solution because
no local minimum of $E_m^{\rm l.a.}$ exists. Applying Derek's
theorem the chiral angle can be shown to collapse when $G$ is
equated to its local approximation. Nevertheless from the small
contribution of the Dirac sea to the total energy we deduce that the
fermion determinant only provides a small repulsive force. The local
approximation furthermore suggests that this repulsive force decreases
with increasing constituent quark mass $m$. This force should cause
$G$ to deviate from the local approximation according to the equation
of motion (\ref{eqng}). This deviation should in turn stablize
$E_{\rm m}$ yielding a solution to the differential equation
(\ref{diffeqn}). Stated otherwise: A significant deviation from the
local approximation is needed to obtain stable self--consistent solutions.

\begin{table}
{}~
\newline
\tcaption{Contributions to the energy for self--consistent solution
as functions of the constituent quark mass $m$. All numbers are in MeV.}
{}~
\newline
\centerline{
\begin{tabular}{c|c c c}
$m$ & ~~400~~ & ~~500~~ & ~~600~~ \\
\hline
$\epsilon_{\rm val}$ & -354  & -436  & -531 \\
$E^{\rm det}[G]$ & 242  & 170 & 131 \\
$E_m$ & 165 & 189 & 157 \\
$E_{\rm tot}$ & 407 & 359 & 288 \\
\end{tabular}}
\end{table}

These features are actually reflected by the self--consistent solutions
to eqns (\ref{diffeqn},\ref{eqng}). For the specific case of $m=400$MeV
the profile functions are plotted in figure 5.1. Technically we have
produced this solution by diagonalizing the unrotated Dirac Hamiltonian
(\ref{hthrho}). The resulting eigenfunctions have been transformed by
$\Omega (\Theta)$ and substituted into the equation of motion
for $G$ (\ref{eqng}). We have then verified that the results are indeed
independent of $\Theta$. {\it I.e.} the calculation of the fermion
determinant has been repeated using $\Theta+\delta\Theta$ without
altering the results. $\delta\Theta$ has been chosen to satisfy the
appropriate boundary conditions. The polarization of the Dirac sea
indeed causes a sufficient deviation of $G$ from its local
approximation to yield stable solutions. On the other hand only a
small repulsive effect is obtained resulting in profile functions which
have a very small spatial extension. At $r\approx1/4$fm the chiral angle
has already dropped to $1/10$ of its value at $r=0$. In the same way
the total collapse is avoided, the upper component of the valence quark
wave--function becomes non--vanishing. Simultaneously the
corresponding energy eigenvalue increases from $-m$ as shown in
table 5.2. {\it I.e.} particle characteristics are admixed.
Furthermore table 5.2 shows that the part of the energy which is
due to the fermion determinant is larger than for the local
approximation ({\it cf.} table 5.1). This suggests that this part of
the energy does not vanish for field configurations where the profiles
are collapsing to a $\delta$--like shape. In this way the total collapse
is avoided and stable solutions do exist. Nevertheless the valence
quark orbit remains strongly bound and possesses properties commonly
attached to presence of the $a_1$ meson. Thus we conclude that indeed
the chiral invariant elimination of the $a_1$ meson carries over
information which would be lost if this field were simply set to zero.
A further feature which can be asserted to the presence of the
$a_1$ meson is represented by the small extension of the soliton
profiles (see figure 5.1). Previously it has been demonstrated that
a non--vanishing $A_\mu$--field provides a squeeze of the chiral
angle. This can {\it e.g.} be inferred from figure 1 of
ref.\cite{zu94a}. Unfortunately the total energy gets very small
($\sim400$MeV). Hence one has to wonder whether this model can
successfully applied to the description of baryons without
amendments. It is obvious that a strong repulsion is called for.

{}From the phenomenology of the nucleon--nucleon interaction is seems
likely that the inclusion of the isoscalar vector meson $\omega$ may
provide (at least some) repulsion. However, as already remarked, for the
non--perturbative treatment of the NJL soliton a stringent derivation
of the  associated Minkowski space energy functional is not available.
Fortunately the current approach allows one to incorporate an
approximation to the full treatment of the $\omega$--meson. Although
this approximation may be somewhat crude it should at least provide
a reliable answer to the question: Does the valence quark energy
remain negative?

The approximation we are going to consider relies on a power expansion
of the NJL action in the $\omega$--field\footnote{By construction, this
is an analytical functional of $\omega$. Thus the energy is well
defined in both Euclidean and Minkowski spaces.}. The leading order is
just the coupling to the baryon current $N_C\omega_\mu B^\mu$. The
crudity of the approximation consists of ignoring all other terms in
the expansion of the fermion determinant and assuming the leading order
gradient expression for $B^\mu$. The latter then becomes the
topological current
\be
B_\mu(U)=\frac{1}{24\pi^2}\epsilon_{\mu\nu\rho\sigma}{\rm tr}
\left\{\left(U^{\dag} \partial^\nu U\right)
\left(U^{\dag} \partial^\rho U\right)
\left(U^{\dag} \partial^\sigma U\right)\right\}.
\label{btop}
\ee
The mesonic part of the action, ${\cal A}_m$, contains a term quadratic
in $\omega$. To this end the $\omega$--dependent parts of the
Lagrangian collect up to
\be
{\cal L}_\omega=\frac{1}{2G_2}\omega_\mu\omega^\mu
+N_C\omega_\mu B^\mu(U).
\label{lomega}
\ee
This expression does not contain any derivative of the $\omega$--field.
Hence it may be eliminated by its stationary condition yielding
the ordinary $6^{\rm th}$ order (in gradients of the chiral field) term
\be
{\cal L}_6=\frac{-1}{2}\epsilon_6^2 B_\mu(U)B^\mu(U)
\quad {\rm with}\quad \epsilon_6^2=\frac{6\pi^2N_C}
{m^2_\rho\Gamma\left(0,m^2/\Lambda^2\right)}.
\label{lag6}
\ee
Adding $\int d^4x {\cal L}_6$ to the mesonic part of the action
gives for the corresponding part of the static energy
\be
E_{\rm m}=(\ref{emvec})+\frac{\epsilon_6^2}{2\pi^3}\int \frac{dr}{r^2}
\left(\frac{d\Theta(r)}{dr}\right)^2{\rm sin}^4\Theta(r).
\label{eng6}
\ee
In this case a stable minimum of $E_{\rm m}$ exists even if the local
approximation were assumed for $G(r)$. The equation of motion for
the chiral angle now becomes
\be
\left[1+\frac{\epsilon_6^2}{4\pi^4f_\pi^2}
\left(\frac{{\rm sin}\Theta}{r}\right)^4\right]\frac{d^2\Theta}{dr^2}
&=&-\frac{2}{r}\frac{d\Theta}{dr}+\frac{1}{r^2}{\rm sin}2\Theta
+\frac{2}{r}{\rm sin}\Theta\left(G+1-{\rm cos}\Theta\right)
\nonumber \\
&&+\frac{\epsilon_6^2}{2\pi^4f_\pi^2}\frac{d\Theta}{dr}
\frac{{\rm sin}^3\Theta}{r^4}\left(\frac{{\rm sin}\Theta}{r}
-\frac{d\Theta}{dr}{\rm cos}\Theta\right).
\label{diffeqn6}
\ee
The solution to this equation together with (\ref{eqng}) is plotted
in figure 5.2. As expected the chiral angle receives a sizable
extension. This is also reflected by a large mesonic part of the energy
$E_m=1607$MeV for $m=400$MeV. In the same manner the $\rho$--meson
profile gets wider and deviates only slightly from its local
approximation. As the profiles get extended the valence quark energy
tends toward $-m$ and regains its antiparticle character which was
already observed in the local approximation. We conjecture that this
property is common to all models which permit a sizable extension of
$G(r)$. Also the contribution of the fermion determinant to the total
energy decreases. {\it E.g.} for $m=400$MeV we find
$E^{\rm det}=132$MeV. Together with $E_m$ this adds up to
$E_{\rm tot}=1739$MeV.

Let us finally compare the results obtained in the present model (where
the axial vector degree of freedom is eliminated in a chirally invariant
way) with other treatments of the isovector (axial) vector mesons in
the context of the NJL model. The numbers for the associated energies
are displayed in table 5.3.

\begin{table}
{}~
\newline
\tcaption{Contributions to the energy for self--consistent solution
in various treatments of the NJL model. Those meson fields which
are allowed to be space dependent are indicated. The constituent
quark mass $m=$400MeV is common. All numbers are in MeV.}
{}~
\newline
\centerline{
\begin{tabular}{c|c c c}
 & $\pi-\rho$ \cite{al90} & $\pi-\rho-a_1$ \cite{al92} & This model \\
\hline
$\epsilon_{\rm val}$ & 313  & -222  & -351 \\
$E^{\rm det}$ & 711  & 543 & 240 \\
$E_m$ & 149 & 393 & 175 \\
$E_{\rm tot}$ & 861 & 937 & 415 \\
\end{tabular}}
\end{table}

It is obvious that either setting the original axial vector degree
of freedom to zero ($A_\mu=0$) or the chirally rotated one
($\tilde A_\mu=0$) amounts to totally different approaches to the
soliton sector of the NJL model. While the first one yields a
positive valence quark energy (for a reasonable choice of parameters),
the orbit is much more strongly bound in the second approach. This
certainly represents an effect due to the axial vector field
$A_\mu\ne0$. Unfortunately the total energy of the chirally rotated
treatment cannot be assigned to any of the other treatments due
to its smallness.

\bigskip
\stepcounter{chapter}
\leftline{\large\it 6. Conclusions}
\bigskip

We have investigated the role of chiral transformations for the
evaluation of fermion determinants. When these transformations
are topologically trivial they provide a useful tool to evaluate
the chiral anomaly. Furthermore they can be used to demonstrate the
equivalence between the hidden gauge and massive Yang--Mills approaches
to vector mesons. As a further application of the chiral rotation we
have shown that the axial vector degree of freedom can be eliminated
without violating chiral symmetry. A generalization to the case when
chiral fields have a topological charge different from zero is not
straightforward. Even though the special transformation we have been
considering is unitary its topological character prevents the
eigenvalues and --vectors of the original Dirac Hamiltonian to be
regained from the rotated Hamiltonian unless the boundary conditions
are transformed to the topologically non--trivial sector accordingly.
Furthermore we have observed that the stationary conditions to
the static energy functional in the topologically distinct sectors
are ${\underline {\rm not}}$ related by the transformation of
the equation of motion. The boundary conditions for the chiral field
obtained from the stationary condition have been found being invariant
under the chiral rotation only when the basis quark fields are taken
from the topological sector associated with the chiral transformation.
Diagonalization of the rotated Dirac Hamiltonian in this basis can
be reformulated into a problem where the induced vector fields
(\ref{indvec}) belong to the topologically trivial sector
({\it cf.} eqn (\ref{hphi})). In order to diagonalize the resulting
operator (\ref{hphi}) the standard basis \cite{ka84,we92} may be
employed. These techniques have been applied to the case when the
chirally rotated axial vector degree of freedom is absent. We have seen
that in this case soliton solutions do exist and that these are totally
different from those which are obtained in the NJL model with the
original axial vector field being neglected. In particular, the chirally
rotated vector meson contains an important information inherited from
the original axial vector meson: A negative valence quark energy.
Unfortunately, the chirally rotated approach as it stands cannot be
considered to be realistic. This is due to the instability of the
mesonic part of the energy in the local approximation for the vector
meson field. We have, however, made plausible that the incorporation of
the $\omega$--meson will provide an approach suitable to describe the
physics of baryons.

Furthermore we have seen that for field configurations wherein the rotated
vector fields possess a reasonable spatial extension leads to a valence
quark energy being of the amount $-m$. {\it I.e.} the valence quark
has (almost) joined the negative Dirac sea; a picture which underlies
Skyrme type models. Thus those amendments of the Skyrme model which
drop the chirally rotated axial vector meson field \cite{ja88,me88,sch89}
gain strong support from the investigations in a microscopic quark
model presented in this paper.

Let us finally point to a possible application of the techniques
developed in section 4 and 5 for a different area of physics: The quark
spectrum in the background field of an instanton field configuration.
This subject has gained some interest recently for the study of the
partition function of QCD\cite{le92}. Assuming temporal gauge, $A_0=0$,
the evaluation of this spectrum can be reduced to an eigenvalue problem
for the Dirac Hamiltonian
\be
\mbox{\boldmath $\alpha$}\cdot\mbox{\boldmath $p$}-
\mbox{\boldmath $\alpha$}\cdot\mbox{\boldmath $A$}
(\mbox{\boldmath $x$},x_4)+\beta m_0
\label{instanton}
\ee
with the time coordinate $x_4$ acting as a parameter. Here
$A_i=V^{\dag}\partial_iV$ is a pure but singular gauge configuration.
$V$ may be chosen in hedgehog form relating color to coordinate space.
An explicit expression is
{\it e.g.} given in eqn (16.50) of ref.\cite{ch84}. The similarity
between the one--particle operators (\ref{instanton}) and
(\ref{hrvec}) is apparent\footnote{Note, however, that $\mbox
{\boldmath $A$}$ contains parity violating pieces.}. In order
to diagonalize (\ref{instanton}) the singularity carried by
$\mbox{\boldmath $A$}$ has to be removed by a topologically
non--trivial transformation. Since this transformation is different
for various time slices a shifting of the quark levels as shown in
figure 4.1 may occur along the path $-\infty<x_4<+\infty$.

\vskip1cm

\nonumsection{Acknowledgement}
We would like to thank U. Z\"uckert for helpful
contributions in the early stages of this work.

\vskip1cm
\appendix
\stepcounter{chapter}
\leftline{\large\it Appendix: Diagonlization of the Dirac Hamiltonian}
\bigskip

Technically the discretized eigenvalues $\epsilon_\mu$ of the Dirac
Hamiltonian ${\cal H}$ (\ref{stham},\ref{direig},\ref{hthrho}) are
obtained by restricting the space $R_3$ to a spherical cavity of
radius $D$ and demanding certain boundary conditions at $r=D$.
Eventually the continuum limit $D\rightarrow\infty$ has to be
considered.  In order to discuss pertinent boundary conditions it is
necessary to describe the structure of the eigenstates of ${\cal H}$.
Due to the special form of the hedgehog {\it ansatz} the Dirac
Hamiltonian commutes with the grand spin operator
\be
{\bf G}={\bf J}+\frac{{\mbox{\boldmath $\tau$}}}{2}
=\mbox{\boldmath $l$}+
\frac{{\mbox{\boldmath $\sigma$}}}{2}+
\frac{{\mbox{\boldmath $\tau$}}}{2}
\label{gspin}
\ee
where ${\bf J}$ labels the total spin and $\mbox{\boldmath $l$}$
the orbital angular momentum. $\mbox{\boldmath $\tau$}/2$ and
$\mbox{\boldmath $\sigma$}/2$ denote the isospin and spin operators,
respectively. The eigenstates of ${\cal H}$ are then as well eigenstates
of ${\bf G}$. The latter are constructed by first coupling spin
and orbital angular momentum to the total spin which is subsequently
coupled with the isospin to the grand spin \cite{ka84}. The
resulting states are denoted by $|ljGM\rangle$ with $M$ being the
projection of ${\bf G}$. These states obey the selection rules
\be
j=\cases{G+1/2, & $l=\cases{G+1 &\cr G &}$ \cr
& \cr G-1/2, & $l=\cases{G &\cr G-1 &}$}.
\label{gstates}
\ee
The Dirac Hamiltonian furthermore commutes with the parity operator.
Thus the eigenstates of ${\cal H}$ with different parity and/or grand spin
decouple. The coordinate space representation of the eigenstates
$|\mu\rangle$ is finally given by
\be
\Psi_\mu^{(G,+)}=
\pmatrix{ig_\mu^{(G,+;1)}(r)|GG+\frac{1}{2}GM\rangle \cr
f_\mu^{(G,+;1)}(r) |G+1G+\frac{1}{2}GM\rangle \cr} +
\pmatrix{ig_\mu^{(G,+;2)}(r)|GG-\frac{1}{2}GM\rangle \cr
-f_\mu^{(G,+;2)}(r) |G-1G-\frac{1}{2}GM\rangle \cr}
\label{psipos} \\ \nonumber \\
\Psi_\mu^{(G,-)}=
\pmatrix{ig_\mu^{(G,-;1)}(r)|G+1G+\frac{1}{2}GM\rangle \cr
-f_\mu^{(G,-;1)}(r) |GG+\frac{1}{2}GM\rangle \cr} +
\pmatrix{ig_\mu^{(G,-;2)}(r)|G-1G-\frac{1}{2}GM\rangle \cr
f_\mu^{(G,-;2)}(r) |GG-\frac{1}{2}GM\rangle \cr}.
\label{psineg}
\ee
The second superscript labels the intrinsic parity $\Pi_{\rm intr}$
which enters the parity eigenvalue via $\Pi=(-1)^G\times\Pi_{\rm intr}$.
In the absence of the soliton ({\it i.e.} $\Theta=0$) the radial
functions $g_\mu^{(G,+;1)}(r),\ f_\mu^{(G,+;1)}(r),$ etc. are given by
spherical Bessel functions. {\it E.g.}
\be
g_\mu^{(G,+;1)}(r)=N_k\sqrt{1+m/E}\ j_G(kr),\quad
f_\mu^{(G,+;1)}(r)=N_k{\rm sgn}(E)\sqrt{1-m/E}\ j_{G+1}(kr)
\label{freesol}
\ee
and all other radial functions vanishing represents a solution to
${\cal H}(\Theta=0)$ with the energy eigenvalues $E=\pm\sqrt{k^2+m^2}$
and parity $(-1)^G$. $N_k$ is a normalization constant.

Two distinct sets of boundary conditions have been considered in the
literature. Originally Kahana and Ripka \cite{ka84} proposed to
discretize the momenta by enforcing those components of the Dirac
spinors to vanish at the boundary which possess identical grand spin
and orbital angular momentum, {\it i.e.}
\be
g_\mu^{(G,+;1)}(D)=g_\mu^{(G,+;2)}(D)=
f_\mu^{(G,-;1)}(D)=f_\mu^{(G,-;2)}(D)=0.
\label{bc1}
\ee
This boundary condition has the advantage that for a given grand
spin channel $G$ only one set of basis momenta $\{k_{nG}\}$ is
involved. These $k_{nG}$ make the $G^{\rm th}$ Bessel function vanish
at the boundary ($j_G(k_{nG}D)=0$). However, this boundary condition
has (among others) the disadvantage that the matrix elements of
flavor generators, like $\tau_3$ are not diagonal in momentum
space. If the matrix elements of the flavor generators are not
diagonal in the momenta a finite moment of inertia will result even
in the absence of a chiral field \cite{re89}. Stated otherwise, in this
case the boundary conditions violate the flavor symmetry. This problem
can be cured \cite{we92} by changing the boundary conditions for the
states with $\Pi_{\rm intr}=-1$
\be
g_\mu^{(G,+;1)}(D)=g_\mu^{(G,+;2)}(D)=
g_\mu^{(G,-;1)}(D)=g_\mu^{(G,-;2)}(D)=0
\label{bc2}
\ee
{\it i.e.} the upper components of the Dirac spinors always vanish
at the boundary. The diagonalization of the Dirac Hamiltonian
(\ref{stham}) with the condition (\ref{bc2}) is technically less
feasible since it involves three sets of basis momenta
$\{k_{nG-1}\},\{k_{nG+1}\}$ and $\{k_{nG}\}$ for a given grand
spin channel. In table A.1 we compare some properties of the
two boundary conditions (\ref{bc1}) and (\ref{bc2}) in the case
when no soliton is present. The first four quantities appearing
in that table show up in various equations of motion when {\it e.g.}
also (axial--) vector mesons are included \cite{sch93b,zu94a}. In case such
a quantity is non--zero the vacuum gives a spurious contribution
to the associated equation of motion. For an iterative solution to
the equations of motion this spurious contribution has to be
subtracted \cite{zu94b}. It should be noted, however, that the relations
listed in table A.1 are all satisfied for both boundary conditions in the
continuum limit $D\rightarrow\infty$.

\begin{table}
{}~
\newline
\tcaption{Properties of the two boundary conditions (\ref{bc1})
and (\ref{bc2}) in the baryon number zero sector. $f(r)$
represents an arbitrary radial function.}
{}~
\vskip0.5cm
\centerline{
\begin{tabular}{lcc}
Quantity& Condition (\ref{bc1})
& Condition  (\ref{bc2}) \\
\hline
$\sum_\mu\Psi_\mu^{\dag} \beta\gamma_5
i{\mbox{\boldmath $\tau$}} \cdot\hat{\mbox{\boldmath $r$}}\Psi_\mu
{\rm erfc}\left(\left|\frac{\epsilon_\mu}{\Lambda}\right|\right)
{\rm sgn}(\epsilon_\mu)=0$
& yes  & yes  \\
\hline
$\sum_\mu\Psi_\mu^{\dag} {\mbox{\boldmath $\alpha$}} \cdot
\left({\mbox{\boldmath $\tau$}}\times
\hat{\mbox{\boldmath $r$}}\right) \Psi_\mu
{\rm erfc}\left(\left|\frac{\epsilon_\mu}{\Lambda}\right|\right)
{\rm sgn}(\epsilon_\mu)=0$
& no & yes  \\
\hline
$\sum_\mu\Psi_\mu^{\dag} {\mbox{\boldmath $\alpha$}} \cdot
{\mbox{\boldmath $\tau$}} \Psi_\mu
{\rm erfc}\left(\left|\frac{\epsilon_\mu}{\Lambda}\right|\right)
{\rm sgn}(\epsilon_\mu)=0$
& yes  & yes  \\
\hline
$\sum_\mu\Psi_\mu^{\dag} {\mbox{\boldmath $\alpha$}} \cdot
\hat{\mbox{\boldmath $r$}}
{\mbox{\boldmath $\tau$}}\cdot
\hat{\mbox{\boldmath $r$}} \Psi_\mu
{\rm erfc}\left(\left|\frac{\epsilon_\mu}{\Lambda}\right|\right)
{\rm sgn}(\epsilon_\mu)=0$
& yes  & yes  \\
\hline
$\tr \left(\beta f(r)\right)=0 $ & yes  & no \\
\hline
$\tr \left({\mbox{\boldmath $\gamma$}} \cdot
{\mbox{\boldmath $\tau$}} f(r)\right)=0 $ & yes  & yes  \\
\hline
$\langle\mu|\tau_i|\nu\rangle=0$ for $k_\mu\ne k_\nu$
& no & yes  \\
\end{tabular}}
\end{table}

So far the discussion of the boundary conditions has only effected
the point $r=D$. In the context of the local chiral rotation it is
equally important to consider the wave--functions at $r=0$. As already
mentioned the solutions to the Dirac equation (\ref{h0}) are given
by spherical Bessel functions in the free case, $\Theta=0$. Except
of $j_0$ these vanish at the origin. In the case $\Theta\ne0$ we
adopt the boundary conditions $\Theta(0)=-n\pi$ thus no singularity
appears in the Dirac Hamiltonian (\ref{h0}) at $r=0$. Therefore
the radial parts of the quark wave--functions may be expressed as
linear combinations of the solutions to the free Dirac Hamiltonian.
{\it E.g.}
\be
g_\mu^{(G,+;1)}(r)=\sum_k V_{\mu k}[\Theta]N_k\sqrt{1+m/E_{kG}}\
j_G(k_{kG}r), \nonumber \\
f_\mu^{(G,+;1)}(r)=\sum_k V_{\mu k}[\Theta]N_k{\rm sgn}(E_{kG})
\sqrt{1-m/E_{kG}}\ j_{G+1}(k_{kG}r)
\label{bescom}
\ee
where the eigenvectors $V_{\mu k}[\Theta]$ are obtained by
diagonalizing the Dirac Hamiltonian in the free basis. It should be
stressed that the use of the free basis is only applicable because
the point singularity hidden in ${\mbox{\boldmath $\tau$}} \cdot
\hat{\mbox{\boldmath $r$}}$ has disappeared. If singularities show up
for certain field configurations the basis for diagonalizing ${\cal H}$
has to be altered. This has been the central issue of section 4.
There we have described that the eigenvalues and --vectors
can be obtained by adopting a basis which is related by a local
chiral transform to the one described in eqn (\ref{freesol}) together
with the appropriate boundary conditions ({\it cf}. eqn (\ref{bc1})
or (\ref{bc2})). Here we will construct an alternative basis for
the $G=0$ channel by explicitly transforming the boundary conditions.
This has the advantage that there is no need to introduce the auxiliary
field $\phi$ as in eqns (\ref{topbasis} and \ref{ansatzphi}).
We therefore consider the application of the rotation (\ref{globalrot})
at $r=0$ to the eigenstates. In the $0^+$ channel this leads to
\be
\Omega(r=0)\Psi_\mu^{(0,+)}(r=0)=
\pmatrix{-if_\mu^{(0,+;1)}(r=0)|0\frac{1}{2}00\rangle \cr
g_\mu^{(0,+;1)}(r=0) |1\frac{1}{2}00\rangle \cr}.
\label{psi0rot}
\ee
It is then obvious that a pertinent basis is given by
\be
N_k\pmatrix {-i\ {\rm sgn}(E)\sqrt{1-E/m}\
j_1(kr)|0\frac{1}{2}00\rangle \cr
\sqrt{1+E/m}\ j_0(kr)|1\frac{1}{2}00\rangle \cr}.
\label{basis0p}
\ee
This is, of course, no longer a solution to the free Dirac equation.
Moreover, a single component of this spinor does not even solve the
free Klein Gordan equation. At $r=D$ the chiral rotation equals unity.
Thus we demand the discretization condition $j_1(kD)=0$ according
to eqn (\ref{bc1}) {\it i.e.} the momenta are taken from the set
$\{k_{n1}\}$. With this basis we have succeeded in eliminating
the singularities at $r=0$ and keeping track of the boundary conditions
at $r=D$. We are thus enabled to numerically diagonalize ${\cal H}_R$.
When comparing with the eigenvalues of ${\cal H}$ we again encounter a
form of the ``infinite hotel story": one state is missing in the negative
part of the spectrum while an additional shows up in the positive part.
The missing state turns out to be the one at the upper end of the
negative Dirac sea, {\it i.e.} $E\approx-m$. Then it is important to
note that in addition to the states with finite $k$, the basis
(\ref{basis0p}) together with the boundary condition $j_1(kD)=0$ also
allows for the ``constant state" with $k=0$. In the continuum limit
($D\rightarrow\infty$) this state is absent. Including, however, this
state for finite $D$ in the process of diagonalizing ${\cal H}_R$
finally renders the missing state. This is not surprising since
application of the inverse chiral rotation $\Omega^{\dag} (r=0)$ onto
this ``constant state" leads to an eigenstate of the free Dirac
Hamiltonian with the eigenvalue $-m$. It should be remarked that for
the free unrotated problem a ``constant state" with eigenvalue
$-m$ is only allowed in the $G^\Pi=1^-$ channel. Although
$\Omega(\Theta)$ does commute with the grand spin operator its
topological character mixes various grand spin channels via the
boundary conditions.

Accordingly the diagonalization of ${\cal H}_R$ in the $G^\Pi=0^-$ channel
demands the basis states
\be
N_k\pmatrix {i\ {\rm sgn}(E)\sqrt{1-E/m}\
j_0(kr)|1\frac{1}{2}00\rangle \cr
\sqrt{1+E/m}\ j_1(kr)|0\frac{1}{2}00\rangle \cr}
\label{basis0m}
\ee
with the boundary condition $j_1(kD)=0$ in order to be compatible with
the Kahana--Ripka \cite{ka84} diagonalization of ${\cal H}$. The
additional ``constant state" needed here corresponds to a state with
eigenvalue $+m$ of the free unrotated Hamiltonian\footnote{The additional
``constant states" thus do not alter the trace of the Hamiltonian.}.

We have finally been able to diagonalize the chirally rotated
Hamiltonian ${\cal H}_R$ in the $G=0$ sector by very tricky means. It
should also be kept in mind that there are now additional states at the
upper (from $0^+$) and lower (from $0^-$) ends of the spectrum which
do not possess ``counterstates" in the $G=0$ part of the spectrum
in the unrotated problem. For the dynamics of the problem they are of
no importance because their contribution to physical quantities is
damped by the regularization. However, their existence reflects the
topological character of the chiral rotation.

In the other channels {\it i.e.} $G\ge1$ we have unfortunately not
been able to construct a set of basis states which rendered the
eigenvalues of the unrotated Hamiltonian along the approach
described above. In the $G=0$ sector we have already seen that a
``global rotation" $-i\mbox{\boldmath $\tau$}\cdot
\hat{\mbox{\boldmath $r$}}\gamma_5$ is needed for the basis states
in order to accommodate the boundary conditions at $r=0$. Furthermore
a mixture of different grand spin channels appears via the boundary
condition since this ``global rotation" deviates from unity at $r=D$.

\newpage

\centerline{\large \bf FIGURE CAPTIONS}
\vspace{1cm}

\noindent
Fig.\ 4.1

\noindent
A schematic plot of the spectrum of the rotated
Hamiltonian ${\cal H}_{nR}=\Omega(n\Theta){\cal H}
\Omega^{\dag} (n\Theta)$.

\noindent
Fig.\ 4.2

\noindent
Comparison of the self--consistent profile in the unrotated
formulation (dashed line) and the solution to eqn (\ref{eqmrot})
(solid line).

\noindent
Fig.\ 5.1

\noindent
The meson profiles (left) and valence quark
wave--function (right) of the self--consistent solution to eqns
(\ref{diffeqn},\ref{eqng}). Also shown is the local approximation to
the vector meson field $G^{\rm l.a.}={\rm cos}\Theta-1$. Here the
constituent quark mass $m=$400MeV is assumed.

\noindent
Fig.\ 5.2

\noindent
Same as figure 5.1 with the $6^{\rm th}$--order term
(\ref{eng6}) included. The upper component of the valence quark
wave--function is negligibly small.

\end{document}